\documentclass[12pt]{article}
\usepackage[body={17.5cm, 22cm},right=2cm]{geometry}
\usepackage{amssymb,graphicx}
\usepackage{amsmath}
\usepackage{epsfig}
\usepackage{hyperref} 

\def\e{{\rm e}}

\def\d{\partial}
\def\l{\left(}
\def\r{\right)}

\newcommand{\be}{\begin{equation}}
\newcommand{\ee}{\end{equation}}
\newcommand{\bea}{\begin{eqnarray}}
\newcommand{\eea}{\end{eqnarray}}
\newcommand{\bg}{\begin{gather}}
\newcommand{\eg}{\end{gather}}
\newcommand{\bseq}{\begin{subequations}}
\newcommand{\eseq}{\end{subequations}}

\newcommand{\Tr}{{\rm Tr}}

\begin{document}
\begin{titlepage}
\begin{center}
{\LARGE\bf  String Axiverse}\\
\vspace{0.5cm}
{ \large
Asimina Arvanitaki$^{a,b}$,
Savas Dimopoulos$^c$, Sergei~Dubovsky$^{c,d}$,\\ 
\vspace{.3cm}
 Nemanja Kaloper$^e$,  and John March-Russell$^f$}\\
\vspace{.45cm}
{\small  \textit{  $^{\rm a}$ Berkeley Center for Theoretical Physics, University of California, Berkeley, CA, 94720}}\\ 
\vspace{.1cm}
{\small  \textit{  $^{\rm b}$ Theoretical Physics Group, Lawrence Berkeley National Laboratory, Berkeley, CA, 94720 }}\\ 
\vspace{.1cm}
{\small  \textit{  $^{\rm c}$ Department of Physics, Stanford University, Stanford, CA 94305, USA }}\\ 
\vspace{.1cm}
{\small  \textit{  $^{\rm d}$
Institute for Nuclear Research of the Russian Academy of Sciences, 
        60th October Anniversary Prospect, 7a, 117312 Moscow, Russia}}\\
\vspace{.1cm}
{\small  \textit{  $^{\rm e}$
Department of Physics, University of California, Davis, CA 95616, USA
}}\\
\vspace{.1cm}
{\small  \textit{  $^{\rm f}$
Rudolf Peierls Centre for Theoretical Physics, University of Oxford, 1 Keble Road, Oxford, UK
}}
\end{center}
\begin{center}
\begin{abstract}

String theory suggests the simultaneous presence of many ultralight axions, possibly populating each decade of mass down to the Hubble  scale $10^{-33}$eV. Conversely the presence of such a plenitude of axions (an ``axiverse") would be evidence for string theory, since it arises due to the topological complexity of the extra-dimensional manifold and is ad hoc in a  theory with just the four familiar dimensions. We investigate how several upcoming astrophysical experiments will be observationally exploring the possible existence of such axions over a vast mass range from $ 10^{-33}$eV to $ 10^{-10}$eV. Axions with masses between $ 10^{-33}$eV to $ 10^{-28}$eV can cause a rotation of the CMB polarization that is constant throughout the sky.
The predicted rotation angle is independent of the scale of inflation and the axion decay constant, and is of order $\alpha \sim1/137$ --within reach of the just launched Planck satellite.
Axions in the mass range $ 10^{-28}$eV to $ 10^{-18}$eV
give rise to multiple steps in the matter power spectrum, providing us with a snapshot of the axiverse that will be probed by galaxy surveys--such as BOSS, and 21 cm line tomography.
Axions in the mass range $10^{-22}$eV to $ 10^{-10}$eV can affect the dynamics and gravitational wave  emission of rapidly rotating astrophysical black holes through the Penrose superradiance process. When the axion Compton wavelength is of order of the black hole size, the axions develop ``superradiant" atomic bound states around the black hole  ``nucleus". Their occupation number grows exponentially by extracting rotational energy and angular momentum from the ergosphere, culminating in a rotating Bose-Einstein axion condensate emitting gravitational waves. 
For black holes lighter than $\sim10^7$ solar masses accretion cannot replenish the spin of the black hole, creating mass gaps in the spectrum of rapidly rotating black holes that diagnose the presence of destabilizing axions. In particular,  the highly rotating black hole in the X-ray binary LMC X-1 implies an upper limit on the  decay constant of the QCD axion $f_a \lesssim 2\times 10^{17}$GeV, much below the Planck mass. This reach can be improved down to the grand unification scale
  $f_a \lesssim 2\times 10^{16}$GeV, by observing smaller stellar mass black holes.

\end{abstract}
\end{center}
\end{titlepage}

\begin{flushright}
~\\
{{\rm The Principle of Plenitude:} \it ``This best of all possible worlds}\\
{\it will contain all possibilities, with our finite experience of }\\
{\it  eternity giving no reason to dispute nature's perfection."}\\
~\\
\hfil{\it {\rm Gottfried Leibniz,} ``Theodicee".}
~\\
\end{flushright}

\tableofcontents

\section{A Plenitude of Axions}
\subsection{The QCD Axion}
The Standard Model QCD action can be appended by the CP-violating topological interaction \cite{'tHooft:1976up}
\be
\label{thetaaction}
S_\theta={\theta\over 32\pi^2}\int d^4x\epsilon^{\mu\nu\lambda\rho}\Tr \,G_{\mu\nu}G_{\lambda\rho} \, .
\ee
This term is a total derivative and does not contribute to the classical field equations. However, it affects physics at the quantum level due to the existence of topologically non-trivial field configurations.  Under the shift $\theta\to\theta+2\pi$ the action (\ref{thetaaction}) changes by $2\pi$, indicating that
 $\theta$ is a periodic parameter, with a period equal to $2\pi$. In the presence of fermions, as a consequence of the chiral anomaly, the actual physical parameter is not $\theta$ itself, but $\bar\theta$---the sum of $\theta$ and the overall
phase of the quark mass matrix
\be
\label{thetabar}
\bar\theta = \theta + \arg\det m_q \; .
\ee
Stringent bounds on the neutron electric dipole moment, $d_n<2.9\cdot 10^{-26}e$\,cm \cite{Baker:2006ts}, imply that $\bar\theta$, if non-zero, should be tiny $\bar\theta\lesssim 10^{-10}$.  This is the strong CP problem---one of the most tantalizing hints for physics beyond the SM. It shares some important features with other more severe fine-tuning problems of the SM---the cosmological constant and the weak hierarchy problems.  Just as its more famous cousins, 
the strong CP problem requires an extreme fine-tuning of apparently unrelated quantities, $\theta$ and $\arg\det m_q$, to produce a tiny $\bar\theta$ as required by observations.  Moreover because of the observed CP-violation
in the quark sector no symmetry of the SM is restored as $\bar\theta\to 0$.
The important difference with the cosmological constant and hierarchy problems is that the smallness of $\bar\theta$ does not appear to be a necessary prerequisite for the existence of life.  Consequently, unlike these
other mysteries there is no ambiguity on whether the strong CP problem has an anthropic or a dynamical solution.   The smallness of $\bar\theta$ is a clear call for a new dynamical mechanism.

An attractive solution to the strong CP problem is to promote $\theta$ to a dynamical field---the QCD axion $a$. At the classical level the axion action is assumed to be invariant under shifts
$a\to a+const$. In other words, at the classical level the axion is a Nambu--Goldstone boson of a spontaneously broken global symmetry (Peccei--Quinn (PQ) symmetry) \cite{Peccei:1977hh,Weinberg:1977ma,Wilczek:1977pj}. 
This symmetry is preserved in the quantum theory at the perturbative level.  However, the PQ symmetry is anomalous in the presence of topologically non-trivial QCD field configurations, explicitly breaking the shift symmetry and generating a periodic potential for the axion.  The resulting axion vacuum expectation value (vev) then automatically adjusts itself to cancel $\bar\theta$, solving the strong CP problem. 
The physical properties of the QCD axion are to large extent determined by the scale $f_a$ of the PQ symmetry breaking, similar to how the low energy pion interactions are fixed by the pion decay constant $f_\pi$.  In particular the axion mass $m_a$ is given by 
\be
\label{QCDmass}
m_a\approx 6\times 10^{-10} \mbox{eV} \l{10^{16}\mbox{GeV}\over f_a}\r  \;.
\ee
Apart from the coupling to gluons (\ref{thetaaction}) the QCD axion may have similar couplings to other gauge bosons, most notably photons, and derivative 
couplings to fermions. The precise numerical values of these couplings are model dependent, but their overall scale is determined by the axion decay constant $f_a$.
Laboratory searches and astrophysical constraints exclude values of $f_a$ below $\sim10^9$ GeV and, for sufficiently large couplings to photons and, assuming dark matter abundance for the QCD axion, a small region around $10^{11}$ GeV \cite{Amsler:2008zzb}. 

It would be especially interesting to find a QCD axion with $f_a\gg 10^{12}$~GeV. Indeed, in this case for typical initial values of the axion field the relic abundance of axions would be higher than the critical density leading to the overclosure of the Universe \cite{Preskill:1982cy, Dine, Abbott}.  This does not imply that high $f_a$'s are excluded---if the observed (quasi)homogeneous Universe were just a small patch of the inhomogeneous Multiverse, 
then the axion initial conditions may be different from place
to place and life can only develop in rare regions with an atypically small axion density \cite{Linde:1987bx}. 
This anthropic argument is of a particularly mild form---the post-inflationary probability distribution of the axion field in the Multiverse is calculable since the axion properties are entirely encoded in a low-energy effective quantum field theory and the axion is defined in the finite range $(0, 2\pi f_a)$ \cite{Tegmark:2005dy}.  Consequently, finding a high $f_a$ QCD axion would not only explain the smallness of $\bar\theta$,
but also provide strong support to the idea of eternal inflation, which is a natural way to create the
Multiverse, and possibly the idea of the string landscape more generally. 

\subsection{String Theory Axions}
\label{stringaxions}

We see that the axion provides a potentially testable dynamical solution to the strong CP problem and its discovery may have even more profound implications. However, at the effective field theory level it is hard to judge how natural it is to have such a  ``fake" global PQ symmetry which is explicitly broken exclusively by QCD.  Note, that in order not to spoil the solution to the strong CP problem all   
other sources of explicit PQ symmetry breaking should be at least 10 orders of magnitude down with respect to the potential generated by the QCD anomaly, and one may be especially cautious about the viability of such a proposal, given the common lore that global symmetries always get broken by quantum gravitational effects \cite{Kamionkowski:1992mf,Holman:1992us,Kallosh:1995hi}.  This makes it natural to inquire whether axions arise naturally in the most developed quantum theory of gravity---string theory. 

Pseudoscalar fields with axion-like properties generically arise in string theory compactifications as Kaluza--Klein (KK) zero modes of antisymmetric tensor fields \cite{Green:1987mn,Svrcek:2006yi}.  Examples of such fields include, for instance, the Neveu--Schwarz 2-form $B_{2}$, present in all string theories, or
the Ramond--Ramond forms $C_{0,2,4}$ of type IIB theory, or $C_{1,3}$ of type IIA theory.  An interesting property of higher-order antisymmetric tensor fields as opposed to scalar fields (0-forms), is that 
upon compactification they typically give rise to a large number of KK zero modes, determined by the topology of the compact manifold ${\cal M}_6$.  For instance, the number of massless scalar fields resulting from a single two-form $B_{MN}$ or $C_{MN}$ is equal to the number of (homologically non-equivalent) closed two-cycles ${\cal C}_i$ in ${\cal M}_6$. Namely, by taking both of the indices of the two-form along a cycle one gets a massless (pseudo)scalar four-dimensional field corresponding to that cycle.
More formally, the KK expansion for the two-form contains 
\be
B={1\over 2\pi} \sum b^i(x)\omega_i(y)+\dots\;,
\label{expansion}
\ee
where $x$ are non-compact coordinates, $y$ are compact coordinates, and $\omega_i$ form the basis of
closed non-exact two-forms (cohomologies) dual to the cycles ${\cal C}_i$,
 {\it i.e.}, satisfying
\be
 \int_{{\cal C}_i}\omega_j=\delta_{ij}\;.
\ee
A similar argument applies to higher-form fields, for instance the number of (pseudo)scalar zero modes corresponding to $C_4$ is equal to the number of homologically non-equivalent four-cycles.  Of fundamental importance to what follows is the fact that in the vast majority of compactifications the number of cycles is quite large, of order several hundred, even up to $10^5$ \cite{Douglas:2006es}. The basic reason for this is simple combinatorics---it is natural to expect many non-equivalent embeddings of a closed two-surface into a reasonably complicated six-dimensional manifold. For instance, even the simplest Calabi--Yau manifold---the six-torus---has ${(6\times 5)/ 2}=15$ different two-cycles and the same number of four-cycles.
Of course, this example is too simple to give rise to a realistic 
gauge and matter field spectrum. Typically, even the simplest string compactifications incorporating the Standard Model gauge group and chiral fermion content give rise to more than a hundred two-cycles.

The four-dimensional scalar fields resulting from this KK reduction are not only massless, but have zero potential as a consequence of the higher-dimensional gauge invariance of the antisymmetric tensor field action. This invariance guarantees that no potential is generated at any order of perturbation theory. However, antisymmetric tensor fields
have Chern--Simons couplings, which are crucial for the Green--Schwarz mechanism for anomaly cancelation.  Upon KK reduction these terms can result in the axionic
couplings (\ref{thetaaction}) to gauge fields \cite{Witten:1984dg}. In type IIB theory this happens, for instance, for a $C_2$ axion if there is a D5 brane wrapped over the corresponding two-cycle, or for a $C_4$ axion if there is a D7 brane wrapped over the corresponding four-cycle, both of which are natural occurrences in the string landscape.  So we see that string theory has a potential of providing {\it many} particles with the qualitative properties of the QCD axion.

Nevertheless, the above arguments do not suffice to say that the QCD axion is predicted by string theory. First, string axions can be at tree level completely removed from the spectrum of light fields by the presence of fluxes, branes and/or orientifold planes \cite{Douglas:2006es,McAllister:2008hb}. For instance, the DBI action of a D5 brane depends on the NS two-form $B_{MN}$, so that wrapping such a brane around a two-cycle makes the corresponding axion heavy, with mass of order the string scale. Similarly, the RR $C_2$ axion can be lifted by wrapping a NS5 brane over the corresponding two-cycle.  
From now on we will focus exclusively on the axions that escape such tree-level liftings.

Secondly, even if an axion does not become heavy due to these tree-level effects its potential always
acquires non-perturbative contributions from one or more of a variety of sources.  These include, worldsheet instantons \cite{Dine:1986zy}, euclidean D-branes wrapping the cycle \cite{Becker:1995kb}, gravitational instantons\cite{Kallosh:1995hi}, and gauge theory instantons \cite{'tHooft:1976up} if the axion couples to a non-abelian gauge group.  In many cases these corrections can be large enough to ruin the solution of the strong CP problem.

We see, that one cannot really make a case that the QCD axion is predicted by string theory, but rather the requirement for one of the string axions to be responsible for a solution of the strong CP problem puts a restriction on string theory model building.  For instance, it disfavors the possibility that all string moduli are stabilized by a supersymmetry(SUSY)--preserving mechanism, as happens, for instance, in  the KKLT scenario \cite{Kachru:2003aw}.  Indeed, if that were the case axions would receive
large SUSY preserving masses together with their superpartners---saxions. On the other hand if not all of the moduli are stabilized in a supersymmetric way, saxions (and axinos) receive masses after SUSY breaking, while axions, being (pseudo)Nambu--Goldstone bosons, are protected from these contributions. 

Note that SUSY breaking masses for moduli typically come out in the 1~GeV$\div$1~TeV range. 
This gives rise to the infamous cosmological moduli problem---light scalar moduli may overclose the Universe and/or their late decays may destroy successful BBN predictions and introduce non-thermal distortions into the CMB spectrum. The most straightforward way to get around this problem is to assume that the expansion rate during inflation was relatively low, such that the moduli are heavy during inflation and not produced, $H_{infl}\lesssim 0.1$~GeV.  The corresponding inflationary energy scale is still quite high $E_{infl}\sim 10^{8}$~GeV, so that if reheating is efficient the temperature is high enough for successful baryogenesis. Such a cosmological 
scenario will be assumed throughout this paper\footnote{In principle this can be relaxed if there is no low energy SUSY. Note, that still there is a quite restrictive bound on the inflation scale from isocurvature perturbations \cite{Fox:2004kb,Hertzberg:2008wr}, if the PQ symmetry is broken during inflation, which is the natural option for  the string theory axions.}. 

As we will argue now, even though one cannot really predict the existence of the QCD axion in string theory,
the assumption that the QCD axion does exist and is responsible for the smallness of $\bar\theta$, combined with the facts about string theory axions reviewed above, not only puts restrictions on string theory model building, but also strongly suggests a rather predictive scenario with a distinctive set of signatures for upcoming cosmological and astronomical observations.

The point is, as we already stressed, there is no reason for the solution to the strong CP
problem to have an anthropic origin. Consequently, neither the very existence of the QCD axion, nor the extreme smallness of all non-QCD contributions to its potential should be a result of a fine-tuning.  Instead, it has to be a natural dynamical consequence of the properties of the compactification manifold giving rise to the string theory vacuum where we live.  But given that string theory compactifications have a potential of producing hundreds of axions it would be really surprising in such a situation if {\it only} the QCD axion were then light.  Note also that of all axions only one linear combination gets a mass from the explicit breaking due to QCD. 
Consequently, we come to the conclusion that if one takes seriously the QCD axion as a solution to the strong CP problem then in string theory one expects to find {\it many light axions}. As we will see, these axions, if they exist, can be observed in a number of different ways.

The two principal parameters characterizing a general string theory axion are its mass $m$ and decay constant $f_a$. Unlike for the QCD axion these two parameters are not related by (\ref{QCDmass}). What values for these parameters can one expect?  First, it's convenient to parametrize the effective four-dimensional axion Lagrangian
as
\be
\label{genaction}
{\cal L}={f_a^2\over 2}\l \d a\r^2 - \Lambda^4 U(a)\;,
\ee
where $U(\theta)$ is a periodic function with a period equal to $2\pi$. 
The overall scale of the potential $\Lambda^4$ is related to $f_a$ and $m$ through
\be
\label{massreln}
m = \Lambda^2/f_a \;.
\ee
Since it is non-perturbative effects that provide the potential in cases where the axion escapes tree level lifting,
we write the scale in the following form,
\be
\label{nonperturbative}
\Lambda^4=\mu^4\e^{-S}\;,
\ee
where $\mu$ is a UV energy scale.  In general more than one type of non-perturabtive effect
contributes to the potential so $\Lambda^4 U(a)$ should be thought of
\be
\sum_i \Lambda_i^4 U_i(a)
\ee
where each of the dynamical scales $\Lambda_i^4$ has the form (\ref{nonperturbative}).   
If the axion potential arises from the superpotential generated by string instantons, then the UV scale $\mu$ is the geometric mean of the string/Planck scale (which sets the scale of instanton physics in this case) and of the SUSY breaking scale $F_{susy}^{1/2}$, 
\be
\mu^4\sim F_{susy} M_{Pl}^2\;. 
\ee

In string theory constructions one finds, in many cases, that $f_a$ and $S$ are related by \cite{Banks:2003sx,Svrcek:2006yi,Svrcek:2006hf}
\be
\label{fSrelation}
f_a\sim {M_{Pl}\over S}\;.
\ee
This relation has a simple geometrical origin.  Namely, the four dimensional
Planck mass is determined by the typical size $L$ of the compactification manifold as
\be
\label{4dMpl}
M_{Pl}^2\sim g_s^{-2}L^6l_s^{-8}\;,
\ee
where $l_s$ and $g_s$ are the string length and the string coupling. Similarly, the axion decay constant is also determined by these parameters and by the area $A$ of the cycle.
For instance, for an axion coming from the RR two-form in IIB theory integrated on a 2-cycle of area
$A\sim L^2$ one has 
\be
\label{4dfa}
f_a^2\sim L^6l_s^{-4}A^{-2}\sim L^2 l_s^{-4}\;,
\ee
where the two factors of $A^{-1}$ came from the powers of the inverse metric in the kinetic term of the two-form  before KK reduction. The action of the string instanton generating the axion potential is given by the product of the tension of the euclidian D brane that wraps the cycle and the area of the cycle,
\be
S\sim g_s^{-1}l_s^{-2}L^2\;,
\ee
thus giving the relation (\ref{fSrelation}).\footnote{Similarly, if the axion potential is generated by the strong dynamics of a gauge sector coming from a stack of D5 branes wrapping the two-cycle, the gauge instanton
contribution goes as $e^{-S}$ with $S\sim 2\pi/\alpha$ where the inverse gauge coupling at the compactification scale is given by $1/\alpha \sim L^2/g_s l_s^{2}$ so that one again arrives at (\ref{fSrelation}). Clearly, all these relations hold up to order one numerical factors.} The value of $f_a$ can be
significantly lower than the esimate (\ref{fSrelation}) if a compactification manifold is significantly anisotropic, or, particularly, if a large amount of warping is present.
However, in explicit examples $f_a$ never comes out parametrically higher than (\ref{fSrelation}) and it was argued that the inequality
\be
f_a\lesssim {M_{Pl}\over S}
\ee
follows from very basic principles of black hole physics (``gravity as the weakest force" conjecture) \cite{ArkaniHamed:2006dz}. Throughout most of the discussion in this paper we assume that the compactification manifold is not too anisotropic and the amount of warping is limited, so that the relation (\ref{fSrelation}) is a good guide for the possible values of $f_a$.  

In the case of the QCD axion not only QCD non-perturbative effects but also string instantons give
contributions to the potential.  To solve the strong CP problem and not lead to a too large $\bar\theta$, string instanton contributions to the potential must be subdominant by a factor of $10^{10}$ compared to QCD, implying, for intermediate scale supersymmetry breaking, a string instanton action
\be
S\gtrsim 200 \;.
\ee

This discussion then suggests the following scenario for the distribution of $f_a$ and $m$ for different axions. The values of $f_a$ are inversely proportional to the area of the corresponding cycle, so they do not change much from one axion to another.  Given that the compactification is such that $S\gtrsim 200$ for string contributions to the
QCD axion, and no special fine tuning is allowed, {\it all} axion decay constants in this scenario are likely
to be close to the GUT scale $M_{GUT}\simeq 2\times 10^{16}$~GeV.  On the other hand, axion masses are exponentially sensitive to the area of the cycles, so that we expect their values to be homogeneously distributed on a log scale. Given that, as argued above, one can expect several hundred different cycles this suggests that there may be several string axions per decade of energy.  It has also been argued recently that the mixing of axions with vacuum energy cancelling 4-forms in the
Bousso-Polchinski landscape can yield axion mass scannings \cite{kalsor}. In Section 2 we show that there are a variety of observational
windows on such axions.

\subsection{Wilsonian Scanning of the Cosmological Constant}

Interestingly, the same number---of order several hundred cycles---is also suggested by the requirement that there are enough different fluxes to implement the Bousso--Polchinski mechanism for scanning the cosmological constant \cite{Bousso:2000xa,Polchinski:2006gy}.
The famous $10^{500}$ estimate for the number of different  string vacua comes out exactly as a result of having $500$ different fluxes, corresponding to different cycles. At this point one may wonder whether the axion potential itself may be responsible for the scanning of the cosmological constant in the presence of several hundred axions.

Indeed, as explained in Ref.~\cite{ArkaniHamed:2005yv}, in the presence of $N$ decoupled scalar fields with two or more non-degenerate minima each with energies of order $M_{GUT}^4$ there are of order $2^N$ minima that scan the total vacuum energy down to $\sim 2^{-N} M_{GUT}^4$.  In fact, there is no reason
why the minima of the individual fields should all be at the same scale $M_{GUT}^4$. For instance, the opposite extreme is to have at each energy scale a field with several minima, which cancels the vacuum energy at this particular scale. An intuitive way to think about this kind of scanning is to
write the vacuum energy in the binary system. If each field has two minima, the choice of the minima amounts to putting either 0 or 1 at the corresponding digital place.
We will refer to this model of scanning as Wilsonian scanning. 
Of course, in a generic case, instead of one of these two extreme cases of either uniformly distributed scalars or all scalars at the GUT scale, there could be densely populated mass ranges with gaps in between them.  The general condition for successful scanning is that the number of scalars above any energy $E_0$ is larger than $\sim\log(M_{Pl}/E_0)$. 

In order that axions be responsible for such scanning, it is necessary that the axion potential has non-degenerate minima.  If the axion potential is generated by a string instanton it is dominated by a single
contribution (because the $n$-instanton action is $nS$) and the periodic function $U( \theta)$ in (\ref{genaction}) is simply $U(\theta)=1-\cos\theta$ and there is only a single minimum.  On the other hand, if the potential is generated by strong IR gauge theory dynamics there is no reason for a strong suppression of the higher instanton contributions, so $U(\theta)$ is a generic periodic function and it is natural for it to have several non-degenerate minima. A simple toy example of such a situation is when the one- and two-instanton contributions enter with comparable coefficients, $U(\theta)\sim 2 - (\cos\theta+\cos 2\theta)$.

Consequently, if light axions are (partially) responsible for the scanning of the cosmological constant at low energies they should be accompanied by a large number of strongly coupled QCD-like hidden sectors with low confinement scales. This both opens up interesting phenomenology associated to the presence
of this ``dark world" and raises the question of how it managed to escape being observed so far. We will
touch on some of the issues involved in the concluding Section \ref{discussion}.   For now we focus upon the
observational signatures of the light axions that we have argued are generic to string theory once the strong CP problem is solved.
 

\section{Cohomologies from Cosmology}
\label{signatures}

\begin{center}
\begin{figure}[th] 
\includegraphics[width=7in]{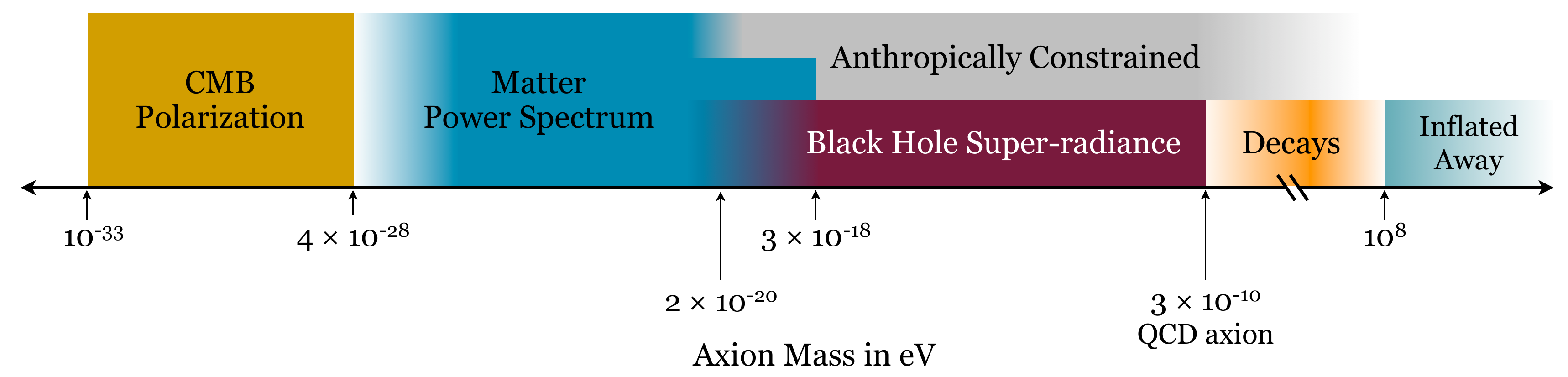}
\caption{{\bf Map of the Axiverse:} The signatures of axions as a function of their mass, assuming $f_a \approx M_{GUT}$ and $H_{inf}\sim 10^8$ eV. We also show the regions for which the axion initial angles are anthropically constrained not to over-close the Universe, and axions diluted away by inflation. For the same value of $f_a$ we give the QCD axion mass. The beginning of the anthropic mass region ($2 \times 10^{-20}$ eV) as well as that of the region probed by density perturbations ($4 \times 10^{-28}$ eV) are blurred as they depend on the details of the axion cosmological evolution (see Section \ref{steps}). $3 \times 10^{-18}$ eV is the ultimate reach of density perturbation measurements with 21 cm line observations. The lower reach from black hole super-radiance is also blurred as it depends on the details of the axion instability evolution (see Section \ref{BHreach}). The region marked as ``Decays", outlines very roughly the mass range within which we expect bounds or signatures from axions decaying to photons, if they couple to $\vec E \cdot \vec B$. We will discuss axion decays in detail in a companion paper.}
\label{bigplot}
\end{figure}
\end{center}

\subsection{Discovering the String Axiverse}

We now turn to the observational consequences of axions lighter than or around the QCD axion mass. For simplicity, we keep $f_a$ fixed at $M_{GUT}$ and $H_{infl}\sim 0.1$ GeV. The initial displacement of axions heavier than $\sim 10^{-20}$ eV has to be tuned in order for them not to overclose the universe and axions heavier than 0.1 GeV have been diluted away by inflation.   The observational consequences of the string axiverse
are outlined in Figure \ref{bigplot}.

We concentrate on three main windows to the axiverse.  First, as discussed in Section \ref{CMBpolarization} axions of masses between $10^{-33}$ eV and $4 \times 10^{-28}$ eV, if they couple to $\vec E \cdot \vec B$, cause a rotation in the polarization of the Cosmic Microwave Background (CMB) by an angle 
of order $10^{-3}$, which is close to the current bounds ($\sim10^{-2}$) \cite{Komatsu:2008hk,Wu:2008qb}.  Future experiments, such as Planck and CMBPol will be able to probe values of the rotation in the CMB polarization down to $10^{-5}$ \cite{Yadav:2009eb}.

Second, as discussed in Section \ref{steps}, axions with masses higher than $10^{-28}$~eV can be a significant component of the dark matter (DM) and  suppress power in small scale density perturbations $\l < 1~\mbox{Mpc}\r$.  This is because the quantum pressure scale originating from the uncertainty principle and below which gravitational collapse is not possible, is proportional to $1/\sqrt{m}$ and thus for these light axions is a cosmologically observable scale. Since the axiverse should contain a plethora of string axions with masses homogeneously distributed on a log scale, the existence of {\it multiple steps} in the small scale perturbation spectrum is a natural expectation.  The amount of the suppression---the step height---is proportional to the fraction of the DM constituted by the particular axion.  Such steps may be detectable with the BOSS \cite{BOSS} and 21 cm line observations \cite{Loeb:2003ya}. 
In particular, the 21 cm line tomography will be sensitive to masses up to $3 \times 10^{-18}$ eV that are well inside the anthropic regime.

Finally, axions of masses between $10^{-22}$ and $10^{-10}$ eV can affect the dynamics of rotating 
 black holes due to the effect of superradiance. When a black hole rotates, a boson with a Compton wavelength comparable to the black hole size creates an exponentially growing bound state with the black hole. This gravitational atom can be de-excited through graviton emission that carries away the black hole's angular momentum. 
 For black hole masses larger than  $\sim10^7~M_{\odot}$, or axion masses smaller than $10^{-18}$ eV, accretion
 may still be efficient enough to support the maximal rotation and 
 sustain a ``Carnot cycle" that turns the black hole into a gravity wave pulsar with possibly detectable signal at future gravity wave experiments. 
 For lighter black holes (heavier axion masses) this effect leads to a spindown of the black hole, resulting in gaps in the mass spectrum of
 rapidly rotating black holes.
 With the quality of data constantly improving, measurements of the spin of stellar mass ($\sim2\div10~M_{\odot}$) black holes will be able to probe also the QCD axion parameter space with $f_a>10^{16}$ GeV, well inside the region where the QCD axion relic abundance 
 is  anthropically constrained. These effects are discussed in Sections 2.4 and 2.5, and in the Appendix.

For axions in the range $\sim 10^{-9}$ to $\sim 10^{8}$ eV, and assuming the axions have couplings to
$\vec E \cdot \vec B$, decays to photons can potentially lead to signatures.  A companion paper will discuss such decays, as well as the physics induced by warping the axion decay constant to scales lower than $M_{GUT}$, and the many dark sectors implied by Wilsonian scanning and/or highly warped throats.

\subsection{Rotation of the CMB Polarization}
\label{CMBpolarization}

Axions much lighter than the QCD axion, when they have an $\vec E \cdot \vec B$ coupling to electromagnetism (EM), change the polarization of the CMB photons if they start oscillating anytime between recombination and today. These axions cannot couple to QCD, as 4d gauge coupling unification implies, otherwise they would get large contributions to their masses.  A coupling to QCD, however, can be easily avoided in the framework of orbifold GUTs \cite{Kawamura:2000ev,Altarelli:2001qj,Hall:2001pg,Hebecker:2001wq}.  An example of such a theory is a scenario with one extra dimension where $SU(5)$ is preserved in the bulk and the breaking down to $SU(3)_c \times SU(2)_L\times U(1)_Y$ occurs on the boundary of the extra dimension. The SM gauge couplings are given by
\be
\frac {1} {g_c^2} =\frac{V} {g_5^2} + \frac{1} {h^2_c},~\mbox{with }c=1, 2, 3
\ee
where $V$ is the extra-dimensional volume, $g_5$ is the 5d $SU(5)$ coupling and $h_c$ are the gauge couplings of the $SU(3) \times SU(2)\times U(1)$ brane kinetic terms that are allowed by the brane--localized breaking of $SU(5)$.  When the volume of the extra dimensions is parametrically large (equivalently when the effective 4d coupling of the $SU(5)$ is much smaller than the brane-localized gauge couplings), there is apparent $SU(5$) unification for the SM gauge couplings,
\be
\frac{1} { g_c^2} \approx \frac{V} { g_5^2},~\mbox{with }c=1, 2, 3 \;,
\ee
with corrections that are parametrically of the same size as traditional GUT-scale threshold corrections.
In these scenarios, since gauge coupling unification is not true everywhere in the extra dimensions, axions that have brane localized couplings, naturally couple to $SU(2)_L$ or $U(1)_Y$ without coupling to QCD.

Only a few ``local" axions are able to couple to electromagnetism with full strength in this way.  Most axions resulting from antisymmetric forms are localized on cycles far away from the position of the SM in the full compactification.   On the other hand, axions from cycles intersecting with ours have a kinetic mixing with our axion and could be more weakly coupled to $SU(2)_L$ or $U(1)_Y$.   Specifically, defining 
\be
\gamma_{ij} = \int_M \omega_i \wedge ^*\omega_j \; ,
\ee
where $\omega_i$ is the basis of the closed two-forms dual to the cycles ${\cal C}_i$ as in (\ref{expansion}), the kinetic terms of the four-dimensional axion fields are of the form
\be
\frac{1}{2}\int d^4 x \gamma_{ij} \partial^\mu b_i \partial_\mu b_j  \; .
\ee
The axions also receive a variety of contributions to their potentials leading to spectrum of axion masses $m_i$.
The end result after diagonalization is that the axions from intersecting cycles also acquire a coupling to electromagnetism, which is suppressed, however, by the mixing angle $\theta_{ij}\sim m_i^2/m_j^2$ between
the axions. This can be a significant suppression for widely separated dynamical scales $\Lambda_i$.

For axions that couple only to EM, interactions are summarized by the following Lagrangian
\be
\label{photonaxlagran}
{\cal L}=-{1\over 4}F_{\mu\nu}^2+{1\over 2}(\d_\mu \phi)^2-\Lambda_a^4U({\phi/ f_a})+{C\alpha\over 4\pi f_a}\phi\epsilon^{\mu\nu\lambda\rho}F_{\mu\nu}F_{\lambda\rho}\;.
\ee
Here  a dimensionless constant $C$ is of order one for an axion that coupled to the photon directly. For instance, in the four-dimensional $SU(5)$ GUT it is equal to $C=4/3$.
When the photon wavelength is short compared to a typical length scale of variation in the axion field, 
the combinations $\vec D\equiv  \l \vec E + {1\over 2} {C\alpha\over \pi f_a} \phi \vec B\r$ and $\vec H \equiv  \l \vec B - {1\over 2} {C\alpha\over \pi f_a} \phi \vec E\r$ satisfy free wave equations \cite{Harari:1992ea}.  Consequently, a region of space with an inhomogeneous axion field becomes optically active \cite{Carroll:1989vb,Harari:1992ea,Lue:1998mq, Pospelov:2008gg}---a linearly polarized freely propagating photon experiences a rotation of the polarization plane by an angle $\Delta\beta$ equal to
\be
\label{rotation}
\Delta\beta={C\alpha\over 2 \pi f_a}\int d\tau \dot \phi\;,
\ee
where $\tau$ is the time along the photon trajectory.  Due to frequent Compton scatterings off electrons CMB photons are not polarized before recombination, so the integration in (\ref{rotation}) goes from the time of recombination, $\tau_{rec}$, till today, $\tau_0$.\footnote{Photons also experience Compton scattering during reionization. 
 To incorporate this effect one has to solve the appropriate kinetic equations rather than just the Maxwell's equations with an axionic source. As soon as the axion mass is much larger than the Hubble parameter of the Universe at reionization this effect is unimportant and the integral in (\ref{rotation}) is saturated at early times.
Also it affects the spectrum mainly at small $l$'s, where the cosmic variance is large.} As a result one obtains for the rotation angle
\be
\label{rotationresult}
\Delta\beta={C\alpha\over 2 \pi f_a}\l \phi(\tau_{0})-\phi(\tau_{rec})\r\;.
\ee
Note that this result is only valid for sufficiently small axion masses, $m\lesssim \delta t_{rec}^{-1}$, where $ \delta t_{rec}\sim 10\mbox{kpc}$ is the duration of recombination.  For larger masses rapid oscillations of the axion field on the timescale of recombination lead to suppression of the rotation angle by a factor $\e^{-m\delta t_{rec}}$.  The rotation angle is maximum for axion masses smaller than the expansion rate at recombination $H_{rec}\sim \l7\times 10^5\mbox{ yr}\r^{-1}$ and bigger than the current Hubble parameter $H_0\sim \l10^{10}\mbox{ yr}\r^{-1}$, while it is still substantial for masses up to $\sim 10 H_{rec}$. In this regime the axion field has negligible value today, $\phi(t_0)\approx 0$. On the other hand at recombination axion oscillations had not yet started, so that the axion field took its primordial value set during inflation, $|\phi(\tau_{rec})|\sim{\pi f_a / \sqrt{3}}$.  As a result, the rotation angle from axions of mass between $10^{-33}$ eV and $4 \times 10^{-28}$ eV becomes,
\be
\label{bestrotation}
\Delta \beta \sim {C\alpha\over 2 \sqrt{3}}\approx 10^{-3}\;.
\ee
 
If $N$ axions are present in this mass range, the rotation angle gets enhanced by a factor of $\sqrt{N}$. The remarkable feature of this result is that the rotation angle (\ref{bestrotation}) is independent of both the PQ scale $f_a$, and the scale of inflation.  As previously mentioned, we assume that $H_{inf} \sim 0.1$ GeV, so isocurvature fluctuations from the axion field are extremely suppressed. Consequently, the rotation angle $\Delta\beta$ is  constant throughout the sky.  
As a result of this rotation a part of the E-mode polarization gets converted into the B-mode. Therefore, even though, as a consequence of a low inflationary scale, we do not expect
the gravitational wave B-mode signal in the axiverse, B-mode can nevertheless be generated through this effect.
The two sources of the B-mode are easily distinguishable, because axions generate also EB and TB cross-corellations which are forbidden by parity in the standard case. 
 
Even though current limits \cite{Komatsu:2008hk,Wu:2008qb} on the rotation angle are at the level $\Delta \beta \lesssim 2^\circ=3.5\cdot10^{-2}$, {\it i.e.} above the level of the expected signal for one axion, Planck
will improve this limit by an order of magnitude down to $0.1^\circ$ and will be sensitive to (\ref{bestrotation}). Moreover it is expected that CMBPol will be sensitive to a signal as low as $0.005^\circ$ \cite{Yadav:2009eb}.  Consequently, in the future we will be sensitive to axions even somewhat outside the optimal mass range, or to smaller axion couplings to $\vec E \cdot \vec B$, when the signal is suppressed as compared to (\ref{bestrotation}).
 \subsection{Steps in the Power Spectrum}
\label{steps}

A purely gravitational signature of light axions is the presence of step-like features in the matter power spectrum. 
The suppression of the CDM power spectrum at  small scales in the presence of ultra-light scalar fields is well-known \cite{Hu:2000ke,Amendola:2005ad} and is very similar to the suppression due to free-streaming of light massive neutrinos. In this section, we review  the origin of the effect and  estimate the relevant scales and the amount of suppression for string theory axions.

The axion field starts oscillating when the expansion rate drops below the axion mass at a time,
\be
\label{Hm}
H(\eta_m)=m\;. 
\ee 
Before $\eta_m$ the axion is frozen at a constant value, while afterwards it starts oscillating and its energy density redshifts away like ordinary CDM,
\be
\label{fastphi}
\phi(\eta)=\phi_0\l{a_m\over a(\eta)}\r^{3/2}\cos m\int_{\eta_m}^\eta a(\eta)d\eta\;.
\ee
A perturbation in the axion field, $\delta \phi$, with comoving momentum $k$ satisfies
\be
\label{perturbationeq}
\delta\ddot{\phi}+2{\dot a\over a}\delta\dot\phi+k^2\delta\phi+m^2a^2\delta\phi-4\dot\phi\dot\Psi+2m^2a^2\phi\Psi=0\;,
\ee
where $\Psi$ is the perturbation of the gravitational potential (here we are working in Newtonian gauge and neglecting the anisotropic stress tensor, so that $\Psi$ uniquely characterizes scalar metric perturbations). At times later than $\eta_m$ (and also such that the physical momentum of the mode is smaller than the mass, c.f. (\ref{mcrossing}))
it is convenient to separate the oscillatory part of the axion field solution by writing
\be
\label{psi}
\delta\phi={1\over 2}\l\e^{i mt}\psi+\e^{-i mt}\psi^*\r\;,
\ee
where $\psi$ is a slowly varying function of space and time. Then the density perturbation in the axion field  averaged over times longer than the period of oscillation is given by
\be
\delta_a\equiv{\delta\rho_a\over \rho_a}=-\Psi+\l{a\over a_m}\r^{3/2}{\psi+\psi^*\over\phi_0}\;,
\ee
and the axion field equation translates in the following equation for $\delta_a$ in the large mass limit
\be
\label{deltaeq}
\ddot\delta_a+{\dot{a}\over a}\dot\delta_a+\l{k^2\over 2ma}\r^2\delta_a=-k^2\Psi+3\ddot\Psi+3{\dot{a}\over a}\dot{\Psi}\;.
\ee 
This equation coincides with that for ordinary CDM apart from the last term on the lhs, which indicates the presence of a momentum dependent sound velocity,
\be
c_s^2={k^2\over 4m^2 a^2}\;.
\ee
This ``quantum pressure" is a manifestation of the uncertainty principle for the axion particles and can be neglected if the physical momentum of a mode is smaller than the Jeans momentum $\sqrt{Hm}$. Assuming that the initial axion field is homogeneous, as implied by our choice of $H_{inf}\sim0.1$ GeV,  $\delta_a$ satisfies adiabatic initial conditions and behaves just as CDM for scales larger than the Jeans scale.

The evolution of a density perturbation of momentum $k$ is ultimately determined by the time $\eta_c$ of horizon crossing,
\be
\label{hcrossing}
{k\over a(\eta_c)}=H(\eta_c)\;, 
\ee
and the time where the physical momentum becomes smaller than the axion mass and the mode is no longer relativistic,
\be
\label{mcrossing}
{k\over a(\eta_r)}=m\;. 
\ee 
Modes with comoving momentum
\be
\label{km}
k<k_m\equiv {m^{1/2}H_0^{1/2}\Omega_m^{1/4}\over z_{eq}^{1/4}}\approx 0.01\mbox{ Mpc}^{-1}\l{m\over 4\times 10^{-28}\mbox{ eV}}\r^{1/2}
\ee
become non-relativistic while still being superhorizon (``long modes"), later the axion starts oscillating and finally the mode enters inside the horizon,
\be
\eta_r<\eta_m<\eta_c\;.
\ee
Note, that in (\ref{km}) we assumed that the axion is heavy enough, such that the beginning of oscillations happens in the radiation dominated era $\eta_m\ll\eta_{eq}$. For these modes the additional term in equation (\ref{deltaeq}) can be neglected and they behave like ordinary CDM.

\begin{figure}[tbp] 
   \begin{center}
   \includegraphics[width=5in]{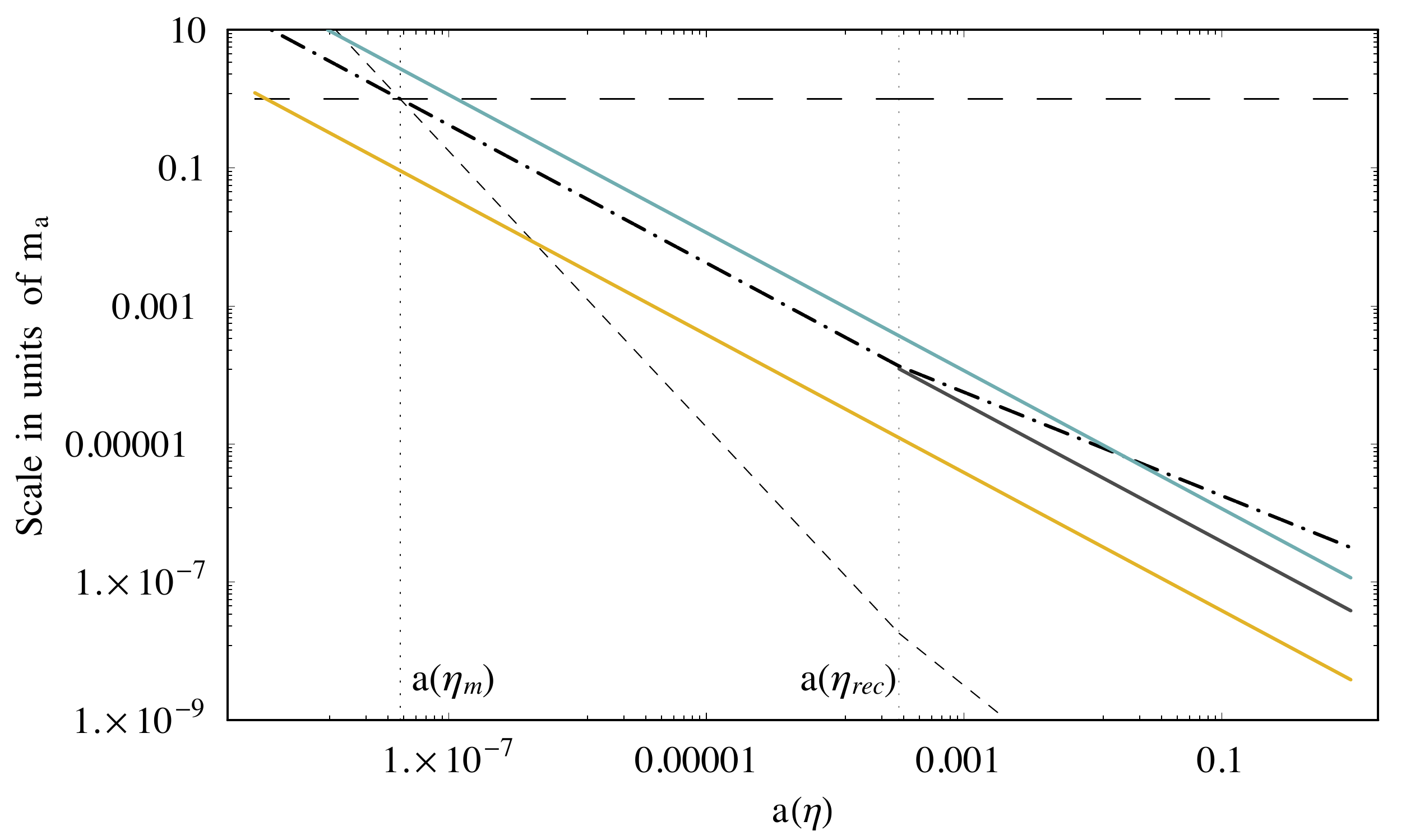} 
   \caption{Evolution of different physical momentum scales as a function of the scale factor.
   The horizontal long-dashed line corresponds to the axion mass. The short-dashed line shows Hubble. The dash-dotted line shows the evolution of
   the Jeans scale. Finally, three solid lines correspond to physical momenta of long and short modes and of the mode with $k=k_m$.}
   \label{history}
   \end{center}
\end{figure}

\begin{figure}[tbp] 
   \begin{center}
   \includegraphics[width=5in]{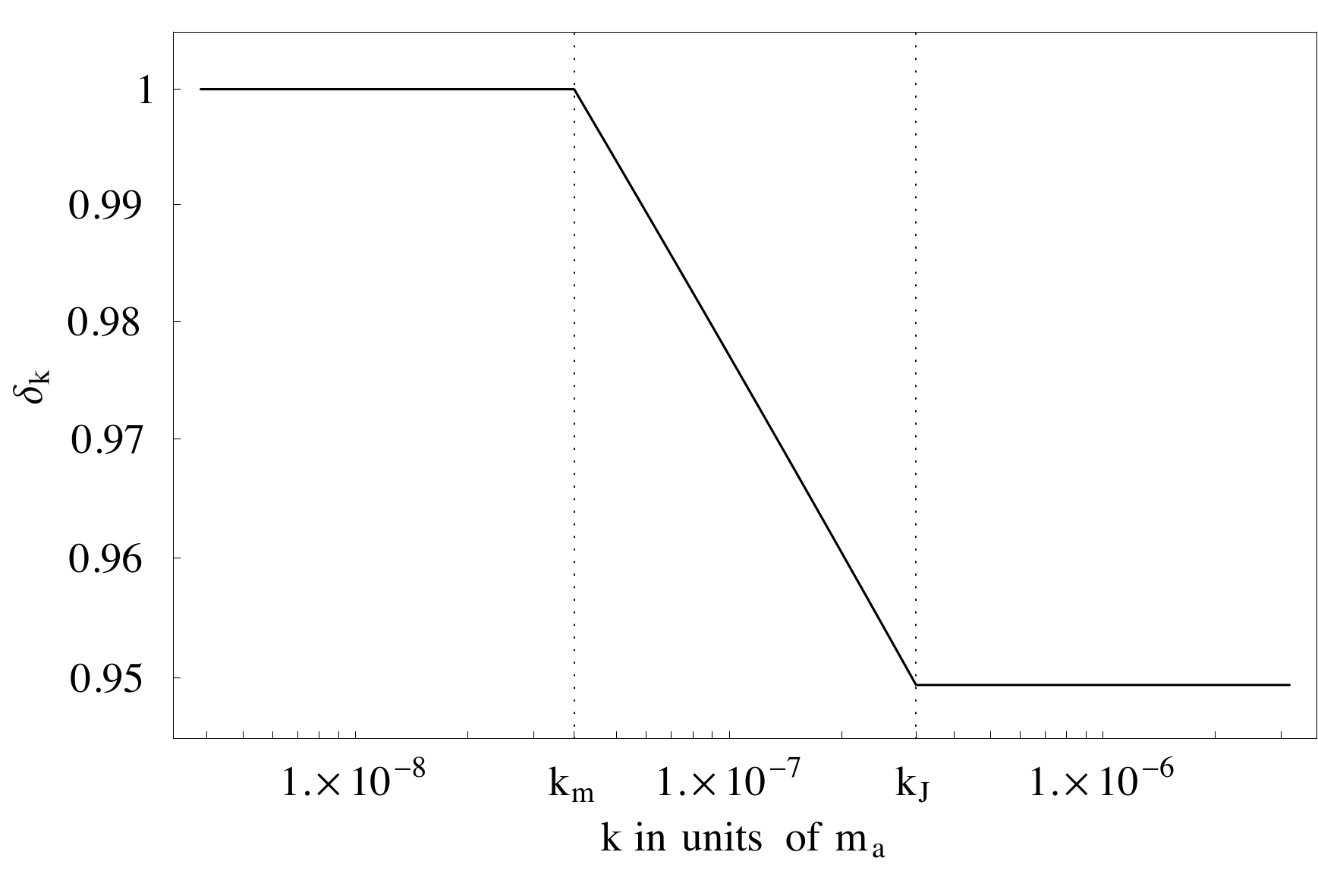} 
   \caption{Suppression of the power spectrum observed today as a function of the comoving monentum. $\delta_k$ has been normalized to the value at large scales and we have assumed ${\Omega_a \over \Omega_m}=0.01$.}
   \label{step}
   \end{center}
\end{figure}

While the presence of an ultra-light axion component does not affect the CDM power spectrum at $k<k_m$, this is no longer true for modes with $k>k_m$ (``short modes''). These modes enter the horizon while still being relativistic, and they keep being relativistic at the moment $\eta_m$, when the homogeneous axion field starts oscillating, so that the characteristic
time-scales are ordered as 
\be
\eta_c<\eta_m<\eta_r\;.
\ee
After $\eta_r$ density perturbations in the axion component can be described by (\ref{deltaeq}). The important difference with the long-mode case is that one cannot neglect the sound speed term in (\ref{deltaeq}) until the even later time $\eta_J$, when the physical momentum of the mode becomes longer than the Jeans momentum
(see Figure~\ref{history})),
\be
\label{jeans}
{k\over a(\eta_J)}=\sqrt{H(\eta_J)m}\;.
\ee
As a result any perturbation in such a mode decays until the moment $\eta_J$.
The Jeans momentum redshifts with the same rate as the physical momentum of the mode during the radiation epoch, so that condition (\ref{jeans}) gets satisfied for short modes only after matter--radiation equality. Since the momentum dependent sound velocity prevents density perturbations from growing, at any moment of time $\eta$ an axion behaves as a smooth homogeneous component of dark matter at physical momenta larger than $\sqrt{H(\eta)m}$. When an ultra-light axion constitutes only a subleading fraction of the dark matter the dominant consequence of such a behavior is a change in the growth rate of the dominant heavy CDM component after matter--radiation equality. Specifically, at this time the Poisson equation for the gravitational potential at scales shorter than the Jeans scale takes the following form
 \be
-k^2\Psi=4\pi Ga^2 \delta\rho_{h}\approx{3\over 2}{\Omega_m-\Omega_a\over\Omega_m}a^2H^2\delta_h\;,
\ee
where the subscript $h$ refers to the dominant heavy component of the dark matter, and in the last equality we neglected the radiation and vacuum energy contributions to the energy density.
Therefore the growth of the density perturbation $\delta_h$ is described by the equation
\be
\label{deltaheq}
\ddot\delta_h+{\dot{a}\over a}\dot{\delta}_h={3\over 2}{\Omega_m-\Omega_a\over\Omega_m}a^2H^2\delta_h\;.
\ee 
During matter dominance the growing solution of this equation evolves as
\be
\label{growth}
\delta_h\propto a^p\;,
\ee
where
\be
p={-1+\sqrt{25-24{\Omega_a\over\Omega_m}}\over 4}\approx 1-{3\Omega_a\over 5\Omega_m}\;.
\ee
As a result, the presence of the ultra-light axion results in a step-like feature suppressing the dark matter power spectrum at small scales. The suppression starts at $k_m$, and at shorter scales the suppression factor grows as
\be
S(z_J)\simeq \l{1+z_{eq}\over 1+z_J}\r^{p-1}\approx 1-{3\Omega_a\over 5\Omega_m}\log {1+z_{eq}\over 1+z_J}
\ee
where $z_{eq}\approx 3200$ is the redshift at matter-radiation equality, and $z_J$ is the redshift corresponding to the moment $\eta_J$ when the physical momentum of the mode becomes longer than the Jeans scale 
$k_J=\sqrt{Hm}$,
\be
1+z_J\approx { \Omega_mm^2H^2_0 \over k^4}\;.
\ee
For modes with momenta higher than $k_J$ at the redshift of observation, $z_o$, the effect saturates; the power spectrum for such modes is suppressed relative to the usual $\Lambda$CDM
case by a factor $S(z_o)$.  Consquently, the presence of a subdominant ultra-light axion component exhibits
itself as a step-like feature in the power spectrum as shown in Figure \ref{step}. The width of the step is almost an order of magnitude in the comoving momenta---the suppression shows up at $k_m\sim(m H_0)^{1/2} (\Omega_m/z_{eq})^{1/4}$ and saturates at around $k_J \sim (m H_0)^{1/2} (\Omega_m)^{1/4}$. 
As one might expect the magnitude of the effect is controlled by the fraction $\Omega_a/\Omega_m$ of the axion density relative to the total CDM density, however there is also an additional logarithmic enhancement,
\be
\label{logfac}
\log{1+z_{eq}\over 1+z_{o}}\approx 8-\log\l{1+z_o}\r\;,
\ee
due to the accumulation of the effect over a long period of time. This discussion assumes $z_o\ll z_{eq}$, at $z_o\sim z_{eq}$ the suppression factor
is simply  $\sim\Omega_a/\Omega_m$.

It is straightforward to estimate the fractional abundance $\Omega_a/\Omega_m$: An axion starts to oscillate when $H\simeq m$, and at this time the total energy density in the Universe is equal to
\be
\rho_{tot} \simeq3M_{Pl}^2{\Lambda_a^4\over f_a^2}\; ,
\ee
while the axion density at the beginning of oscillations is of order $\Lambda_a^4$.
As a result one finds that the axion fraction is of order
\be
\label{axion_fraction}
{\Omega_a\over \Omega_{m}}=P(\theta_i)\cdot{f_a^2\over 3 M_{Pl}^2} \cdot{1+z_m\over 1+z_{eq}}\;,
\ee
where 
\be
z_m\approx {m^{1/2} z_{eq}^{1/4}\over H_0^{1/2}\Omega_m^{1/4}} \approx z_{eq} \l {m\over 4\times 10^{-28}\mbox{ eV}}\r^{1/2}
\ee
is the redshift at $H=m$
and $P(\theta_i)$ is a statistical factor depending on the initial axion angle $\theta_i=\phi/f_a$. 
We see that the size of the steps, which is the product of the fraction (\ref{axion_fraction}) and a logarithmic factor
(\ref{logfac}), depends on three parameters $\theta_i$, $f_a$ and $z_m$ so we now discuss the dependence and typical values we expect for each of them in turn.

\begin{figure}[t!]
\begin{center}
\includegraphics[width=0.49\textwidth]{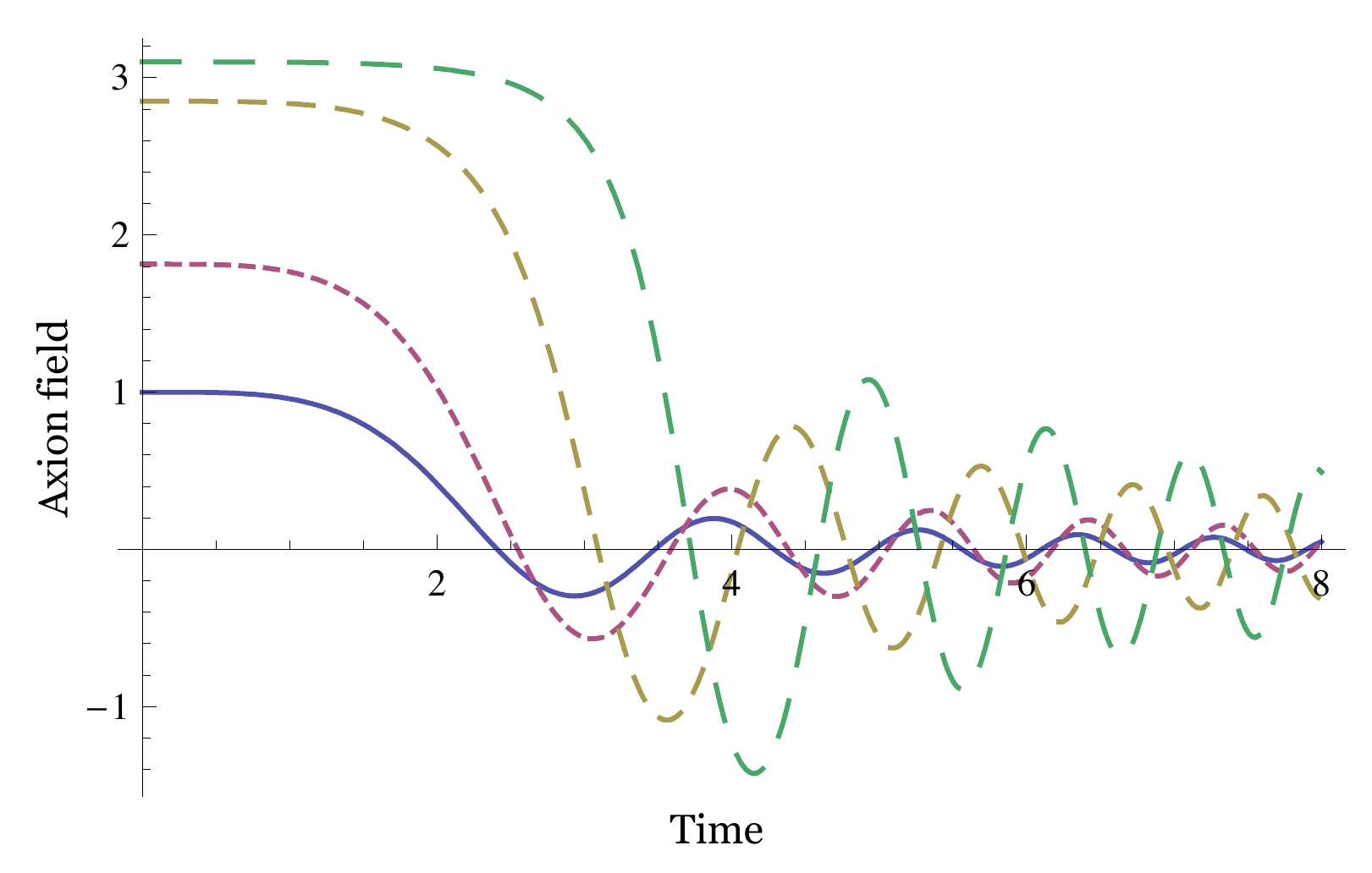}
\includegraphics[width=0.49\textwidth]{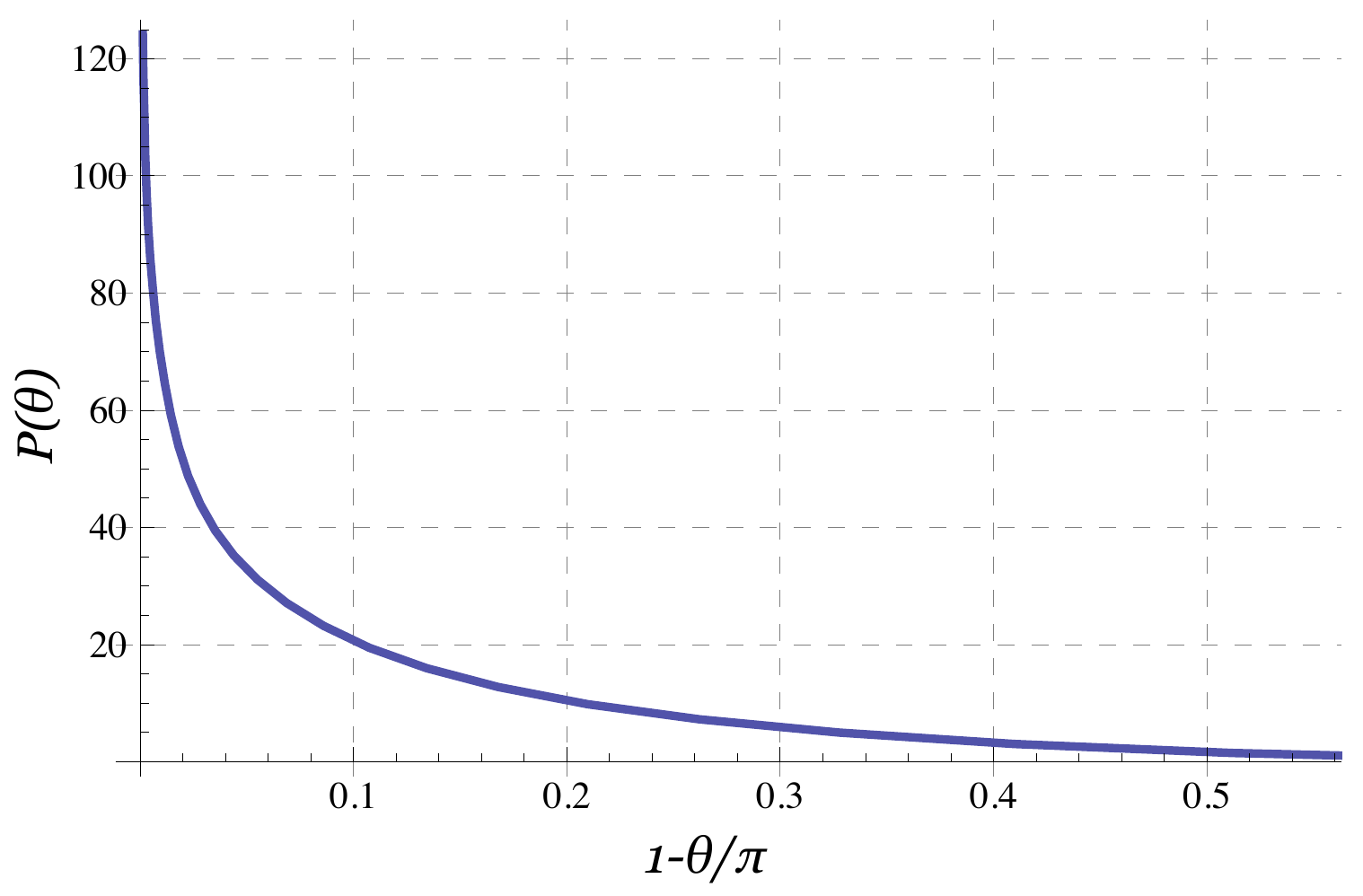}
\end{center}
\caption{
A time evolution of the axion field for different initial conditions (left panel) and an enhancement factor $P(\theta)$ for the axion abundance as a function of a probability for different 
initial axion values (right panel).
}
\label{angledependence}
\end{figure}

We illustrate the dependence on the initial axion angle in Figure~\ref{angledependence} (c.f. \cite{Bae:2008ue}). On the left panel we present solutions for an axion field with different initial conditions in a radiation dominated Universe as a function of a scale factor normalized to one at $H=m$. We see that increasing the initial value of the axion  gives rise to two effects.  First, the resulting axion amplitude is higher just because
the initial condition is higher.  Second, the axion potential becomes flatter close to $\theta=\pi$, and as a result oscillations start later also resulting in the increase of the axion abundance.  We illustrate this effect on the right hand panel of Figure~\ref{angledependence}, where we present a numerical result for the enhancement factor $P(\theta)$ as a function of the probability $(\pi-\theta)/\pi$ to have an initial axion misalignment higher than $\theta$. 

The crucial microscopic parameter that determines a size of the effect is the ratio $f_a^2/(3M_{Pl}^2)$. As discussed in Section \ref{stringaxions}, it is widely believed that this ratio is necessarily smaller than one, and
the relation ${f_a/M_{Pl}}\lesssim {1/S}$ should hold, where for the ultralight axions that we consider here the instanton action should be quite big, $S\gtrsim 200$.  This favors rather small values $f_a^2/(3M_{Pl}^2)\sim10^{-5}$.  On the other hand, these estimates have an order one uncertainty in them, so that values as high as $f_a^2/(3M_{Pl}^2)\sim10^{-4}$ may be not unreasonable.

Finally, the third factor in (\ref{axion_fraction})---the redshift ratio---is determined by the axion mass. Different types of observations will be sensitive to different values of masses.
For instance, CMB measurements are sensitive for the CDM power at scales crossing the horizon
around matter--radiation equality. From the above discussion we see that there is highly unlikely to be an effect at the level significantly higher than $10^{-3}$ at these scales (note also, that in this case $z_o\sim z_{eq}$ so one looses the enhancement of (\ref{logfac})), so the CMB is not a good probe of this effect. 

At the shorter scales the limits on the CDM power spectrum can be obtained from galaxy surveys and Ly-$\alpha$ forest data. The step-like feature in the power spectrum due to ultra-light axion is very similar to the suppression of the small scale power due to a free-streeming neutrino component (c.f. \cite{Amendola:2005ad}), so to estimate the sensitivity of the currently available data 
we can translate the bounds  on the warm-plus-cold dark matter models involving sterile neutrinos (see \cite{Boyarsky:2008xj} for the up-to-date analysis).  The bounds are dominated by the
SDSS  Ly-$\alpha$ data, which are   sensitive for the comoving momenta in the range $k\sim 0.1\div10  $~Mpc$^{-1}$ and correspond to the observation redshift $z_o\sim2\div 4$. For axion masses close to the lower end of the above momentum range this analysis allows admixture
of an axion component at the level $\Omega_a/\Omega_m\sim 0.1$, while for masses corresponding to $k\gtrsim 4$~Mpc$^{-1}$ an order one fraction of  an ultralight axion component  is allowed.
Taking into account that matter--radiation equality corresponds to $k_{eq}\sim 0.01$Mpc$^{-1}$, and that the characteristic comoving momentum $k_m$ scales linearly with $z_m$ we conclude 
from (\ref{axion_fraction}) that these bounds are still not good enough to probe the most interesting region $f_a^2/(3M_{Pl}^2)\sim10^{-5}$.

On the other hand, the Baryon Oscillation Spectroscopic Survey (BOSS---which is a part of SDSS III \cite{BOSS}) will have a sensitivity at the level of few percent to the CDM power spectrum at comoving momenta $k\sim 0.1$Mpc$^{-1}$, which correspond to axion masses around $4\times 10^{-26}$~eV (cf. (\ref{km})). Given that matter--radiation equality corresponds to $k_{eq}\sim 0.01$Mpc$^{-1}$ the redshift ratio gives a factor $\sim 20$ at these scales (as follows from Figure~\ref{angledependence}a) the axion starts oscillating at a redshift at least half that of $z_m$). For these observations $z_o\sim 1$ so the log enhancement (\ref{logfac}) is maximal. 
Therefore, with a reasonable statistical factor, $P(\theta)\sim 20$, BOSS will be able to observe the effect of axions down to $f_a^2/(3M_{Pl}^2)\sim 10^{-5}$. 

In the more distant future, a high precision measurement of axion steps from 21~cm line measurements is possible. Indeed, observations of the 21~cm line have a chance
to probe the power spectrum in the range $k\sim 10^{-2}\div10^3$Mpc$^{-1}$ \cite{Loeb:2003ya}.  These observations correspond to $z_o\sim 30\div 200$, so one looses
slightly the enhancement factor (\ref{logfac}). However, this will be easily compensated by the increase of $z_m$ and by the high accuracy that can be achieved by 21~cm measurements.
Also, unlike for the Ly-$\alpha$ forest, this has an advantage of probing the power spectrum in the linear regime.
To appreciate the precision one  may hope to achieve, it is enough to note that the 21~cm line will provide a total of $N_{21}\sim 10^{16}$ independent modes as opposed to $N_{cmb}\sim10^7$ modes for the CMB.

Note, that axions with lower masses tend to produce more pronounced steps, so 
that at the early stages when only the low $k$ data is available, rather than observing an individual step one may discover an overall running of the spectral index due to a joint effect of several axions
with different masses.

As follows from (\ref{axion_fraction}) an important scale is $k\sim 10^{4\div 5}k_{eq}$---where axions with the corresponding masses are expected to constitute an order one fraction of the CDM for a typical initial misalignment. Axions with heavier masses would overclose the Universe for a typical misalignment, so their initial amplitudes are forced to be fine-tuned to small values in our patch of the Multiverse. On the other hand, there is no reason for initial values of the lighter axions to be fine-tuned, so we expect that axions with masses close to the anthropic boundary, {\it i. e.} of order
\be
\label{axion_mass}
m\sim {H_0\Omega_m^{1/2}z_{eq}^{3/2}\over P(\theta)^2}\l{3 M_{Pl}^2\over f_a^2}\r^2\approx 1.4\times 10^{-20}\mbox{eV}{1\over P(\theta)^2}\l {3 M_{Pl}^2/f_a^2} \over 10^4 \r^2
\ee
to constitute an order one fraction of the CDM.  The 21 cm line measurements can probe part of the anthropic regime, as they are sensitive to masses up to $3 \times 10^{-18}$ eV. The level of the expected signal is uncertain  though, and may depend on the probability measure that determines the fine-tuning of the initial axion amplitudes.

The value of the axion mass in (\ref{axion_mass}) is also interesting as it has been previously suggested \cite{Hu:2000ke} that the problem with excessive small-scale structures predicted in the vanilla $\Lambda$CDM model may be solved if dark matter is composed of a Bose--Einstein condensate of ultra-light particles with masses of $\sim 10^{-22}$~eV. Of course, it may be that the problem will be resolved within conventional $\Lambda$CDM after more precise $N$-body simulations including baryons become available. Still, it is intriguing that a similar mass scale appeared in the present context as the boundary between anthropic and non-anthropic axions, so that one expects an order one fraction of CDM to be composed of axions in this mass range  (recall that the statistical factor of $P(\theta)\sim 10$ has a probability 0.2 for a single axion). Note, that this connection, if true, favors high $f_a^2/(3M_{Pl}^2)$ ratios around $10^{-4}$.  On a positive note this increases the likelihood of seeing light axions with BOSS, but it forces a reconsideration of whether such 
a high ratio is achievable for ultra-light axions in explicit string theory constructions.

\subsection{Extraction of  Black Hole Rotational Energy by Axions }
\label{BHbomb}
Ultralight axions with Compton wavelengths comparable to the size of astrophysical black holes 
can be sought for in observations of rapidly spinning black holes. The key reason for this is the Penrose process, which opens the channel for the black hole spindown. Indeed, the spinning black holes have the so-called ergoregion, a region outside of the event horizon and therefore accessible to external observers, 
inside which the inertial frame dragging due to the black hole rotation is so fast that probes built of
normal causal matter, which can never move faster than light, cannot remain at rest relative to an observer far away. This opens the possibility for energy and spin loss by the black hole \cite{Penrose:1969pc,Christodoulou:1970wf}, which happens when a particle falls into the ergoregion such that it co-rotates with the black hole. If after diving into the ergoregion the projectile splinters up into two fragments, and one falls into the black hole while the other recoils away and out of the ergoregion, the escaping fragment can take out more energy and spin than the original projectile. 

To see this in a bit more detail, consider the geometry of the spinning black hole given by the Kerr solution in Boyer-Lindquist coordinates 
\cite{Wald:1984rg}
\bea
&&ds^2 = - (1-\frac{2R_g r}{\Sigma}) dt^2 - \frac{4R_gar \sin^2\theta}{\Sigma} dt d\phi + \frac{\Sigma}{\Delta} dr^2 + 
\Sigma d\theta^2 + \frac{(r^2+a^2)^2 - a^2 \Delta \sin^2 \theta}{\Sigma} \sin^2\theta d\phi^2 \, , \nonumber\\
&& ~~~~~~~~~~  \Sigma = r^2 + a^2 \cos^2 \theta \, , ~~~~~~~ \Delta = r^2 - 2R_gr + a^2 \, , ~~~~~~~ a = \frac{J}{M},~~~~~~~R_g=G_NM \, , 
\label{kerrsoln}
\eea
and where $M$ and $J$ are black hole's mass and spin, respectively. The roots of $\Sigma = 2R_gr$ define the ergosphere, whose interior is the ergoregion, while the roots of $\Delta = 0$ are the black hole horizons, 
and the outer one, relevant for our discussion, resides at $r_+ = R_g + \sqrt{R_g^2 - a^2}$. The spatial location of the outer horizon never extends past the ergosphere, reaching it only at the poles.
The time at asymptotic infinity is measured by $t$, and represents the clock reading of an inertial Minkowskian observer far from the black hole. The time evolution of a physical system is therefore described by the flow generated by the vector field ${\cal H} =  \partial_t$. Because this vector field is also a Killing vector of the geometry (\ref{kerrsoln}), if we consider an inertial particle of mass $\mu$ in this background, the product of its 4-momentum $p^\mu={dx^\mu}/{d\tau}$   (where $d\tau^2 = -ds^2$) 
 and of the Hamiltonian is conserved, yielding the energy integral of motion,
\be
\label{energy}
E=-{\cal H} p\;.
\ee
Similarly, since (\ref{kerrsoln}) is manifestly axially symmetric, and the rotations around the $z$-axis are generated by the vector field ${\cal J}= \partial_\phi$, the product  
\be
\label{spin}
L= {\cal J}p
\ee
will be the conserved particle's angular momentum in the $z$-direction.

The crucial property of the ergoregion is that particles moving there may have a negative energy. Indeed, inside the ergoregion the $(tt)$-component of the Kerr metric (\ref{kerrsoln})
is negative, so that the Killing vector ${\cal H}$ is space-like. The 4-velocity of any physical observer should be time-like, implying that all observers experience rotation inside the ergosphere, $d\phi/d\tau>0$. The energy (\ref{energy}), being the product of a time-like and a space-like vectors, is not sign-definite, and takes negative values for some of the observers.


This observation is the key to the Penrose process of black hole energy loss. Imagine a projectile falling toward the black hole, initially coming in from afar where $ {\cal H}$
is time-like. Then its total conserved energy must be positive. Allow it to dive beneath the ergosphere, but aim it such that it misses the horizon. Then design a timer on the projectile to set off a fragmentation process that will break the projectile into two parts, but so that one fragment flies off into the black hole along a trajectory on which the conserved energy is negative, ${\cal E} < 0$. Since the total $4$-momentum conservation governs the fragment dynamics, the energy of the escaping fragment will be $\hat E = E - {\cal E} = E + |{\cal E}|$, {\it exceeding} the energy of the initial projectile. Its gain comes at black hole's expense: the conservation of the total energy of the system implies that 
\be
\delta M = - (\hat E - E) = {\cal E} < 0 \, . \label{bhmasschange}
\ee
As a result of this process the rotation of the black holes spins down.
Indeed, the vector field ${\cal G} = {\cal H} + \omega_+ {\cal J}$  is a Killing vector that becomes null on the horizon.
 Here $\omega_+ = {a}/({2R_gr_+})$ is the angular velocity of the horizon.
 For later purposes, we note that as a function of the spin $a$ and the gravitational radius $R_g$, it is given by
\be
\omega_+ = \frac{1}{2R_g} \frac{{a}/{R_g}}{1+\sqrt{1 - (a/R_g)^2}}
\label{spinvelo}
\ee
Note, that in the extremal limit $a/R_g = 1$ it saturates at $R_g\omega_+({\rm max}) = 0.5$.
Now, the product of the  $4$-momentum ${\cal P}$ of the infalling fragment (which is a future directed time-like vector)
and of the future-directed null vector ${\cal G}$ 
 must obey ${\cal G} \cdot {\cal P} < 0$. Using the definition ${\cal G}$ and Eqs. (\ref{energy}) and (\ref{spin}), this yields for the  spin of the infalling particle 
 ${\cal L} < {\cal E}/{\omega_+}$. Consequently,  by total angular momentum conservation, there will be a change of the black hole's spin as well,
\be
\delta J = - (\hat L-L) = {\cal L} < \frac{\cal E}{\omega_+} < 0 \, . \label{bhspinchange}
\ee
Eqs. (\ref{bhmasschange}) and (\ref{bhspinchange}) show that for Penrose fragments, both the
energy ${\cal E}$ and the spin ${\cal L}$ are negative. Combining these two equations we find that the
black hole hair change by 
\be
\delta J < \frac{\delta M}{\omega_+} \, , 
\label{hairs}
\ee
which in fact is an equation that hides in itself the second law of black hole thermodynamics. Indeed, one can compute the horizon area of the spinning black hole (\ref{kerrsoln}) to find 
\be
A = 16\pi R_g^2\frac{1 + \sqrt{1 -  a^2/R_g^2}}{2} \, ,
\label{irrmass}
\ee
such that 
\[
\delta A = 8\pi\frac{a}{ \sqrt{R_g^2-a^2}} \l\frac{\delta M}{\omega_+} - \delta { J}\r\;.
\] 
By Eq. (\ref{hairs}) this yields
\be
\delta A > 0 \, ,
\label{arealaw}
\ee
precisely the area law. In principle, it is possible to fine-tune the fragment trajectories in the Penrose process such that $\delta A = 0$, and so mine out black hole energy stored in the spin in the amount of 
\be
\delta {\cal E} = {M\over \sqrt{2}}\l\sqrt{2}-{\sqrt{1 + \sqrt{1 - a^2/R_g^2}}}\r \, , 
\label{energygain}
\ee
which means the more energy will come out if there was more spin to begin with. The extraction is maximized for (near) extremal black holes, where it reaches $29\%$ of the total mass.
 Note that the energy stored in spin is a very sensitive function 
of $a/R_g$: for example, for  $a/R_g \sim 0.8$, the energy is already significantly smaller,
$\delta {\cal E}/M \sim 0.106$, whereas for a slowly spinning black hole with $a/R_g \sim 0.3$, it is 
$\delta {\cal E}/M \sim 0.0116$. This energy is essentially analogous to the difference of the total `$mc^2$' energy and rest energy $m_0 c^2$ in special relativity, and shows that the black holes with spins $a/R_g>0.9$ are very relativistic, and so fast-spinning, those with $a/R_g \sim 0.8$ or so are analogous to `warm' particles with kinetic energies comparable to the rest mass, while those with smaller spins, e.g. $a/R_g \sim 0.3$ are already `slow'.

The above discussion was presented in terms of point particles, while we are interested in applying the Penrose process to scalar fields---axions.
One way to see that one can use them to extract energy from the black hole is to reverse the logic and note that the inequality (\ref{hairs}) is a direct consequence of the area law 
(\ref{arealaw}), which is nothing but the second law of thermodynamics. Consequently the inequality (\ref{hairs}) should hold independently of what carries the energy and spin into the black hole. Applied to the incoming scalar wave of the form
$
\Psi=e^{-i\omega t+im\phi}f(r,\theta)
$
with a frequency in the range
\be
\label{wrange}
0<\omega<m\omega_+
\ee
this inequality implies that both energy and spin transfer from such a wave into a black hole should be negative \cite{Zeld,Misner:1972kx, Starobinskii}. 
A direct way to see this \cite{Wald:1984rg} is to consider a conserved energy current for the scalar field $\Psi$ given by
\[
P_\mu=-T_{\mu\nu}{\cal H^\nu}=-\d_\mu\Psi\d_t\Psi\;.
\]
Let us now consider a space-time region between two slices of constant time $t$. Conservation of the current $P_\mu$ implies
that the time-averaged energy flux at the infinity is equal to the time-averaged energy flux through the black hole horizon.
The latter is equal to
\be
\langle P_\mu{\cal G}^\mu\rangle=-\langle(\d_t\Psi+\omega_+\d_\phi\Psi)\d_t\Psi\rangle=\omega(\omega-m\omega_+)f^2
\ee  
and is indeed negative in the frequency range satisfying the superradiance condition (\ref{wrange}), so that by scattering 
such waves off the black hole one extracts the energy. Based on this observation Press and Teukolsky \cite{Press:1973zz,Press:Nature} designed a ``black hole bomb"---a spinning black hole 
surrounded by a spherical mirror. Such a device exhibits an exponential classical instability---being confined by the mirror a small initial  scalar field perturbation inside the shell with a frequency in the superradiant 
range (\ref{wrange}) experiences a repeated series of  amplifications by scattering off the black hole, until it extracts all of the black hole's spin (or the mirror blows up).

\begin{figure}[t] 
 \begin{center}
 \includegraphics[width=7in]{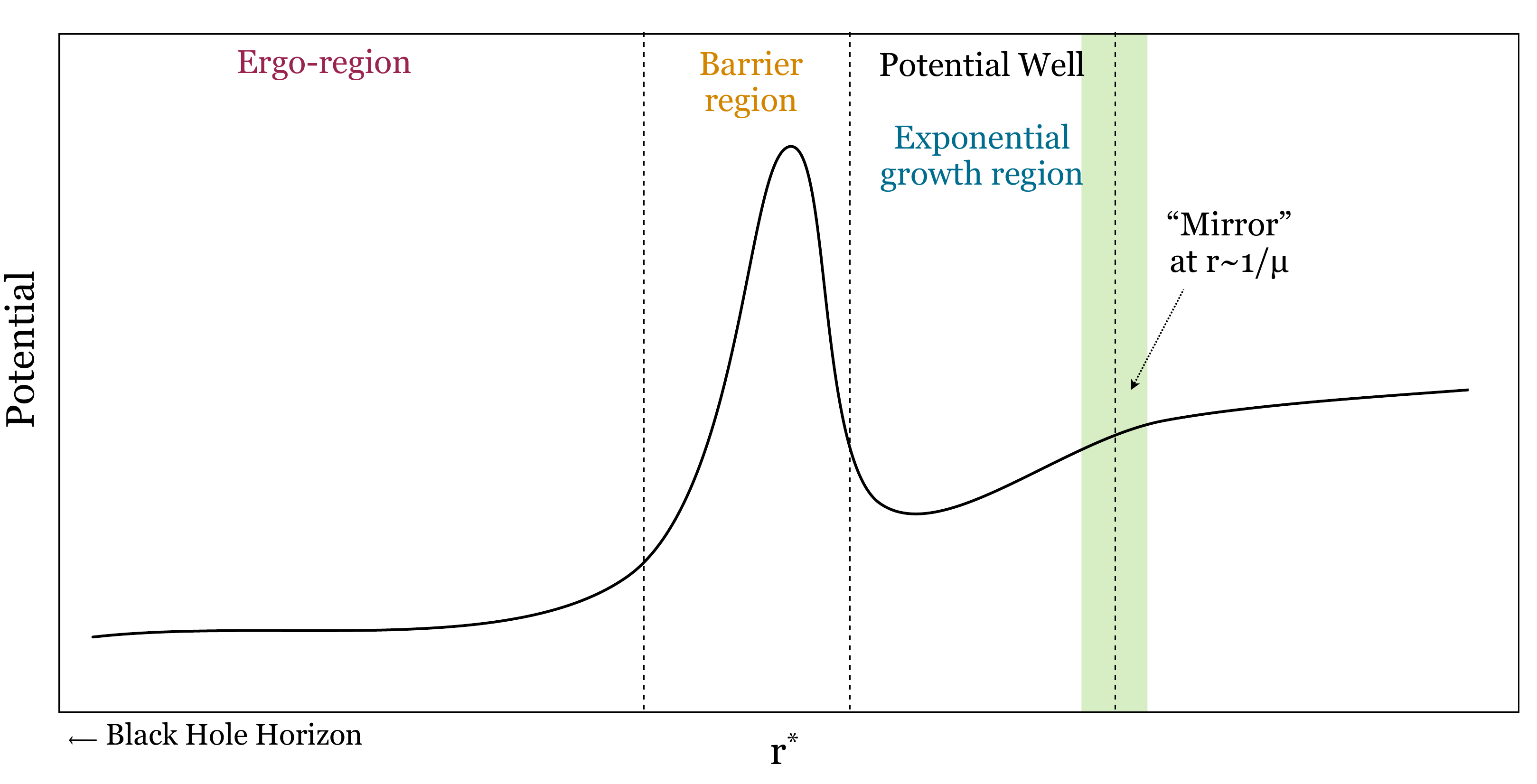}
 \caption{The effective potential of Eq. (\ref{schrodpot}). Depicted are the ergoregion, to the left, with the horizon at  $r^* \rightarrow - \infty$,  the centrifugal barrier (whose height depends on the angular momentum of a mode), the potential well to the right of it, and the asymptotic mass barrier which plays the role of the mirror that reflects the escaping Penrose fragment back. The relevant modes will be the states bound in the potential, and leaking through the barrier towards the horizon.}
 \label{superradiance}

\end{center}

\end{figure}

An extremely interesting observation
 made already in \cite{Damour:1976kh}  is that the Nature itself may provide such a mirror if the field has a non-zero mass. Indeed, unlike massless particles, massive ones can rotate on stable orbits
 around the black hole just like planets around the Sun. Consequently, for a massive scalar 
 field there should be a set of bound states in the Kerr background corresponding to wave-packets rotating along these stable orbits. However, unlike for the point particles, such a wave packet will always have a tail going into the ergosphere region as well. If the wave-packet contains frequencies in the superradiant range (\ref{wrange}) they will be continuously amplified
 and the amplitude of the field will be growing exponentially. In other words, one expects to find an exponentially growing bound states in the spectrum of scalar field perturbations
 in the Kerr background. This intuition was proven to be correct  \cite{Zouros:1979iw,Detweiler:1980uk}. Let us see how unstable modes arise at a more technical level.
 Rewriting $(\Box - \mu^2) \Psi =0$ in 
the Kerr background  (\ref{kerrsoln}) and using the separation of variables $\Psi = R(r) \Theta(\theta) e^{im\phi} e^{-i\omega t}$,
one finds that  $\Theta$ is an oblate spheroidal harmonic, and that $R$ must satisfy 
\be
\Delta \frac{d}{dr} \Bigl( \Delta \frac{dR}{dr} \Bigr) + \Bigl(a^2 m^2 - 4R_g ram\omega + (r^2 +a^2)^2 \omega^2 - \mu^2 r^2 \Delta \Bigr) R = \l\lambda_{ml} + \omega^2 a^2\r\Delta R \, , \label{radeq}
\ee
where $\lambda_{ml}$ are oblate spheroidal eigenvalues, which depend on $m$ and $l$ but in general cannot be written analytically in terms of them. This equation can be cast in a more useful form as a Schr\"odinger equation, by defining the tortoise coordinate $r^*$ according to $dr^* = (r^2 +a^2) dr/\Delta$, and rescaling the wave function to $u = \sqrt{r^2+a^2} R$. The resulting equation is \cite{Zouros:1979iw}
\be
\frac{d^2u}{dr^{*2}} +\l\omega^2 - V(\omega) \r u = 0 \, , \label{schrod}
\ee 
with the potential
\bea
V(\omega) &=&  \frac{\Delta \mu^2}{r^2 + a^2} + \frac{4R_gram\omega - a^2 m^2 + \Delta \l\lambda_{ml}  + (\omega^2-\mu^2)a^2\r}{(r^2+a^2)^2} ~~~~ \nonumber\\ 
&& ~~~~~~~~~~~~~~~~~~~~~~ + \frac{\Delta (3r^2 - 4R_gr + a^2)}{(r^2+a^2)^3} 
- \frac{3\Delta^2 r^2}{(r^2+a^2)^4}  \, .
\label{schrodpot}
\eea
The potential approaches $V \rightarrow V_\infty = \mu^2$ far from  the
hole, when $r^* \rightarrow \infty$, rising out of a potential well where $V < \mu^2$ (see Figure \ref{superradiance}). Note, however, that it rises towards its asymptotic value at infinity already as soon as $r \gtrsim 1/\mu$, and so the effective spatial extent of the potential well is really $\delta r \sim 1/\mu$. This ``mass barrier" plays a role of the mirror which reflects the runaway Penrose fragment back toward the black hole, enabling it to undergo repeated Penrose scatterings and gain more energy and spin from the black hole. On the other side of the well is the centrifugal barrier, peaking at $r^* \simeq R_g$. Beyond the centrifugal barrier is
the ergoregion, where the potential asymptotes to $V \rightarrow V_+ = 2 m \omega_+ \omega - m^2 \omega_+^2$ as $r \rightarrow r_+$ (equivalently, as $r^* \rightarrow - \infty$). 

The  crucial property of the equation (\ref{schrod}) is that it is {\it not} a selfadjoint eigenvalue problem---the potential (\ref{schrodpot}) depends on $\omega$ in an essential way
when $a \ne 0$ (equivalently $\omega_+ \ne 0$). This makes possible the existence of bound states with complex frequencies.
In particular, near the horizon the terms in the parenthesis in 
(\ref{schrod}) combine into $\omega^2 - V_+ = (\omega - m \omega_+)^2$. As a result, the asymptotic 
form of the wave function near the horizon is $\exp(\pm i k_+ r^*)$, where 
\[
k_+=\omega - m \omega_+\;.
\]

The requirement of regularity at the horizon singles out one of these waves, $\exp( -i k_+ r^*)$ \cite{Zouros:1979iw}.
The appropriate boundary condition at infinity, given that we are looking for bound states, is the exponential decay of $u$.
Altogether, the boundary conditions are
\bea
\Psi &\rightarrow& \frac{\Theta_{ml} (\theta)}{\sqrt{2R_gr_+} } e^{-i k_+ r^* + i m \phi - i \omega t} \, , ~~~~ {\rm for} ~~ r^* \rightarrow - \infty \, ,  \label{horizonasymp} \\
\Psi &\rightarrow& \frac{\Theta_{ml} (\theta)}{r}  e^{- \sqrt{\mu^2 - \omega^2} r^* + i m \phi - i \omega t} \, , ~~~~~~~~ {\rm for} ~~ r^* \rightarrow + \infty \, .
\label{waveasymp}
\eea
We see, that bound states with $\mbox{Re } k_+>0$ satisfy ingoing boundary condition at the horizon, {\it i.e.} being unable to escape at the infinity, they still can be sucked into the black hole.
On the other hand, for  $\mbox{Re } k_+<0$, {\it i.e.} for frequencies in the superradiant interval (\ref{wrange}), the boundary condition (\ref{horizonasymp}) describes an {\it outgoing}
flux of particles from the black hole. Of course, this is in a perfect agreement with the previous derivation demonstrating that superradiant modes get amplified as a result of scattering off the black hole. This implies that if the real part of a frequency is smaller than $m\omega_+$, the imaginary part should be positive, indicating the presence of an exponential classical instability.
In the {\it Appendix} we worked out the details of how this happens in a simple toy potential.
The quantitative expressions for the imaginary parts in the actual Kerr background were caculated explicitly in the limit $R_g \mu \gg 1$ in 
\cite{Zouros:1979iw} and in the limit $R_g \mu \ll 1$ \cite{Detweiler:1980uk},  and the numerical 
calculation in a general case was performed recently in \cite{Dolan:2007mj}.
With the notation $\omega = \omega_r + {i}/{\tau}$, the explicit calculations find
\bea
\tau &=& 10^7 e^{1.84 R_g\mu} \, R_g \, , \, ~~~~~~~~~~~~~~~~~ {\rm for} ~ R_g\mu \gg 1 \mbox{ and } a=1 \, , \label{heavy}\\
\tau &=& 24 \Bigl(\frac{a}{R_g}\Bigr)^{-1} \Bigl(R_g \mu)^{-9} \,  R_g \, , ~~~~~~~~ {\rm for} ~ R_g\mu \ll 1 \, . 
\label{instimes}
\eea
For $a<1$ the instability scale at large masses has the same qualitative form (\ref{heavy}), but the  coefficient 1.84 in the exponent grows at smaller $a$.
The following comments are in order  regarding the fastest instability channel. First,  the instability
\begin{itemize}
\item benefits from as large $a$ as possible, being the fastest for near extremal black holes with $a/R_g = 1$.
\end{itemize}
\noindent
Since for the unstable modes $ \omega_r < m \omega_+$, and $\omega_+ \le {1}/(2R_g)$ (see Eq. (\ref{spinvelo})), we find that $\omega_r < {m}/(2R_g)$. Now, the rate of the instability is controlled by the tunneling of such modes through the centrifugal barrier, which is higher at larger $l$. Hence, 
among the available modes the fastest growing one will be the one which tunnels most easily, i.e.
\begin{itemize}
\item has the smallest total angular momentum $l$;
\item has the largest projection on the black hole's axis of rotation, $m=l$.
\end{itemize}
The first condition implies that it has the lowest possible barrier, and the second ensures that it is the highest level in the well, probing the thinnest available section of the barrier. Now, the specific frequencies that can be used depend on the value of $\mu$ as well. If $R_g\mu \gg 1$, then $\mu \gg \omega_+$, and the fastest instability is due to the mode with $m\sim l \sim \mu/\omega_+$, resulting in the  exponential suppression due to the fact that the potential well is very narrow and the centrifugal barrier high and thick. In the opposite limit, $R_g \mu \ll 1$, the fastest instability is in the sector $m=l=1$, which is now available, and is suppressed mostly due to the fact that the fastest mode is spread out through the potential well becoming wider near its top. This results in the power law suppression of the instability rate. 


Numerical results of \cite{Dolan:2007mj} show that the instability peaks around $R_g \mu \sim 0.42$ (for $a/R_g \sim 0.99$), where
\be
\tau_{sr} \sim 0.6 \times 10^7 R_g \, ,
\label{maxinst}
\ee
%
In this regime the instability is due  to the mildly non-relativistic (the real frequency is $\omega_r \sim 0.98 \mu$) $l=m=1$ level.

\subsection{Axionic Sirens and Precision Black Hole Physics}
\label{BHreach}
Black holes are believed to be abundant in our Universe; in particular, there  is  a $4 \times 10^6$ solar mass ($M_\odot$) black hole at the center of our Galaxy. Current black hole candidates are primarily found in X-ray binaries and Active Galactic Nuclei (AGN) and clustered in two mass ranges, 3$M_\odot$ -- 30$M_\odot$ for stellar mass black holes, and  $10^6M_\odot-10^9M_\odot$ for supermassive black holes \cite{McClintock:2009as}. There is
currently an evidence also for intermediate mass black holes \cite{Schodel:2005qm} as well as for  black holes heavier than $10^{10} M_\odot$ \cite{Valtonen:2008aj}. 

In the past few years, a lot of progress has been made in measuring the properties of these black holes including their spin. There is evidence for rotating stellar mass black holes as well as a rotating black hole candidate at $10^7M_\odot$ \cite{McClintock:2009as}. In the next decade X-ray observations, in combination with a better understanding of back hole environments, will solidify and extend black hole spin measurements. In the not-so-near future gravity wave observatories such as LISA and AGIS \cite{Dimopoulos:2008sv} will provide even more precise probes of black hole properties. The natural question arises what   implications the axion superradiance may have for the fate of astrophysical black holes and whether 
observations of astrophysical black holes may provide evidence for ultra-light  axions. 

To address this question the analysis of section~\ref{BHbomb}
has to be extended in several directions.
First, one may worry that the instability may be significantly inhibited or even totally disappear in a realistic astrophysical enviroment. 
Apart from the exponentially unstable superradiant modes Kerr metric also supports bound states in the non-superradiant regime, whose frequencies have {\it negative} imaginary parts. Physically, these bound states get damped rather than amplified by the black hole and eventually get sucked behind the horizon. Some of these modes may have imaginary parts much larger than the maximum value $\tau_{sr}^{-1}\sim 10^{-7}R_g^{-1}$ for the rate
of the superradiance instability \cite{Dolan:2007mj}. Consequently, one may worry that perturbations to the Kerr metric which are always present in a realistic
astrophysical situation may lead to the mixing between superradiant and non-superradiant modes and damp the instability.

Second, if the superradiant instability survives these perturbations and developes, at some point 
the linearized approximation breaks down, and 
the  backreaction of the axion cloud  has to be taken into account to deduce the observational consequences.

\subsubsection{Superradiance in Realistic Enviroments}
Let us start  with checking  that the superradiant instability persists also in a realistic astrophysical enviroment.
We will see that the black hole vicinity is a very clean astrophysical enviroment---as exemplified by no-hair theorems---the black hole itself cleans up the space around its horizon. 

One source of perturbations on the Kerr metric comes from the presence of  accreting matter. A reasonable estimate (in fact, an upper bound in most cases)  for the accretion rate is provided by the Eddington limit \cite{Rees:1984si}---the accretion rate, such that the radiation pressure on free electrons  balances gravity. In the Eddington regime the black hole mass $M$ grows according to 
\be
\label{Eddington}
\dot{M}={M\over \tau_{accr}}\;,
\ee
where the Eddington accretion time $\tau_{accr}$ is equal to 
\be
\label{eddtime}
\tau_{accr}= \frac{{\sigma_{\text{\tiny{Thompson}}}}}{4 \pi G_N m_{\text{\tiny{proton}}}}\approx 4\times 10^8~\mbox{years}.
\ee
From this we can estimate the amount of matter in the vicinity of the black hole horizon as
\be
\delta M\simeq M {R_g\over \tau_{accr}}\approx 4\times 10^{-22} M \l{M\over M_\odot}\r\;.
\ee
The shifts of the imaginary parts one can expect from such a perturbation are of order $ R_g^{-1}\delta M/M$, which is  comfortably smaller than the 
positive imaginary part due to superradiance, if the axion mass is not too far from the optimal regime.

Another source of  potentially dangerous distortions  is a tidal force due to a companion star rotating around the black hole. For instance, stellar mass black holes are observed in X-ray binaries, so such a star is always present.
 The physical (gauge invariant) part of the metric perturbation due to tidal forces caused by a companion star with a gravitational radius $r_g$ at a distance
 $L$
 is of order
 \[
 \delta g_{\mu\nu}\sim R_{\mu\lambda\nu\rho}\Delta x^\lambda \Delta x^\rho\;,
\]
where the Riemann  curvature created by the companion star is of order $R_{\mu\lambda\nu\rho}\sim r_g/L^3$ and  $\Delta x\sim R_g$ is the size of the axion cloud.
Consequently, the tidal force correction to the imaginary part of the superradiant modes is at the level
\be
\label{tidalcorrection}
\delta\omega\sim {r_g R_g^2\over L^3}R_g^{-1}\;.
\ee
To get an idea of how big this correction is let us estimate it for the X-ray binary LMC X-1,  supposedly harboring a  10$M_\odot$ black hole with spin ${a/R_g}\approx0.91$ \cite{Gou:2009ks} (the motivation for this choice is that, as we discuss in Section~\ref{axion_bhbounds},  at the moment LMC X-1 provides the best reach for the QCD axion). The companion star in the LMC X-1 is quite heavy---around 30~$M_\odot$ \cite{Orosz:2008kk}, and its orbital period is $\sim 3.9$ days, which corresponds to the distance $L\sim 2.5\times 10^{12}$~cm. Altogether, this gives rise to a tiny  frequency shift of order $\sim 10^{-18}R_g^{-1}$, which is not dangerous for the superradiance.

As an example of a galactic black hole let us consider Sgr A*--- a $\sim 4\times 10^6 M_\odot$ black hole at the center of the Milky Way. For a number of years  Keplerian orbits of around 30  stars around Sgr A* were monitored \cite{Gillessen:2008qv}, with the fastest period being equal to 15 yr, which 
corresponds to the distance of order $10^4 R_g$, again being safe for superradiance instability. It follows from (\ref{tidalcorrection}) 
that for relatively light galaxies, $M\lesssim 10^7 M_\odot$, a compact stellar size object (such as a neutron star, or a black hole) within $\sim 10 R_g$ from
the horizon may be dangerous for the superradiance instability. However, such  occurances are likely to be quite rare, as follows from the 
estimates for the rate of extremal mass ratio inspirals (EMRI), $\sim few\times100$~Gyr$^{-1}$ \cite{Hopman:2009gd} and are likely not to last long, as the objects
get swallowed by the black hole. Given that the time (\ref{maxinst}) to build up the axion cloud is only $\sim 10^{4}$~yr even for the largest ($10^{10}M_\odot$) black holes, EMRI's effect on the superradiance is also negligible.

Another potential source for the superradiance shutdown, could be  the presence of a magnetic field around the black hole if the axion has a coupling to EM
\be
\label{axphot}
{C \alpha\over 4\pi f_a}\phi\epsilon^{\mu\nu\lambda\rho}F_{\mu\nu}F_{\lambda\rho},
\ee
where C is a constant that is ${4 \over 3}$ in $SU(5)$ GUTs. As the axion instability develops around the black hole, there will be conversion of axions to photons due to the presence of a strong magnetic field and energy will be carried away from the axion field. A strong enough magnetic field
may lead to a total dissipation of the axion cloud.

In the presence of non-zero magnetic $B$ and axion fields the coupling (\ref{axphot}) gives rise to the source term in the Maxwell's equations
\[
\d_\mu F^{\mu\nu}={C\alpha\over 2\pi f_a}\epsilon^{\mu\rho\lambda\nu}F_{\mu\rho}\d_\lambda\phi\sim{C\alpha\over \pi f_a}\mu B\phi\;,
\] 
where we took into account that the axion field oscillates with a frequency of order $m$.
This source produces photons with an energy density of order
\[
\rho_\gamma\sim\l{C \alpha\over \pi f_a}\r^2B^2\phi^2
\]
that take away the energy from the axion field with a characteristic time scale
\[
\tau_\gamma\sim \l{\pi f_a\over C \alpha}\r^2{\mu \over 3B^2}\; .
\]
In order for the superradiance instability to be effective $\tau_\gamma$ should be longer than the characteristic superradiance time $\tau_{sr}$ (\ref{maxinst}) or, equivalently,  
\be
\label{Bless}
B\lesssim 2\cdot 10^{-4}{\pi\over C \alpha}\Lambda^2\;,
\ee
where $\Lambda=\sqrt{mf_a}$, assuming $\mu \sim(G_N M)^{-1}$.

For the QCD axion the bound (\ref{Bless}) corresponds to a magnetic field of order $5\cdot 10^{16}$~G, which is much larger as compared to what the black hole accretion disk can support (Eq. (\ref{Bbh}), \cite{Rees:1984si})
\be
B\sim 4 \times 10^8 G \l{M\over M_\odot}\r^{-1/2} \, .
\label{Bbh}
\ee
Hence the superradiance instability is definitely present, and the magnetic field does not affect the range of $f_a$'s probed. 
Similarly, (\ref{Bbh}) implies that (\ref{Bless}) is satisfied for supermassive galactic black holes as well for $f_a\sim M_{GUT}$.

However, magnetic fields may shut down the superradiance for  $f_a \ll M_{GUT}$. An example is the ultralight axion which can affect supernovae luminosities by photon-axion mixing \cite{ckt}, 
and which operates at the mass scale $\mu \sim few \times 10^{-16} {\rm eV}$ with the effective coupling
${C\alpha}/({4\pi f_a}) \sim (4 \times 10^{11}{\rm  GeV})^{-1}$. This field could be in the regime of the fastest instability of the smallest supermassive black holes on the record, with $M \sim 10^5 M_{\odot}$, 
which are able to support  magnetic fields $B \sim 10^6~{\rm G}$, while the critical $B$-field value for them, by Eq. (\ref{Bless}), is $\sim  10^2~{\rm  G}$. Thus the instability may be turned off in this case.
\subsubsection{The Fate of  the Axionic Instability}
Let us discuss now what happens with superradiance at late stages of the instability development. As always with a linear instability, at a certain
point the backreaction of the axion cloud has to be taken into account and the problem becomes non-linear. Naively, one may think
that at this stage the problem becomes very complicated and hardly tractable. However, the important property of the superradiant instability is that it is always very slow---the instability time scale is at least seven orders of magnitude longer than a natural dynamical time-scale of the system $R_g$, see
(\ref{maxinst}). As a result, second order effects have a chance to compete with the superradiance instability when the axion cloud is still a small perturbation to 
the black hole, and the whole process can be under control.
Still, as we will see, a black hole surrounded by an axionic cloud is an extremely rich dynamical system, 
and our purpose here is just to provide a basic intuition of what one can expect from it. A dedicated quantitative analysis of 
different regimes will appear in a separate publication \cite{inprogress}.

\begin{figure}[tbp] 
 \begin{center}
 \includegraphics[width=5in]{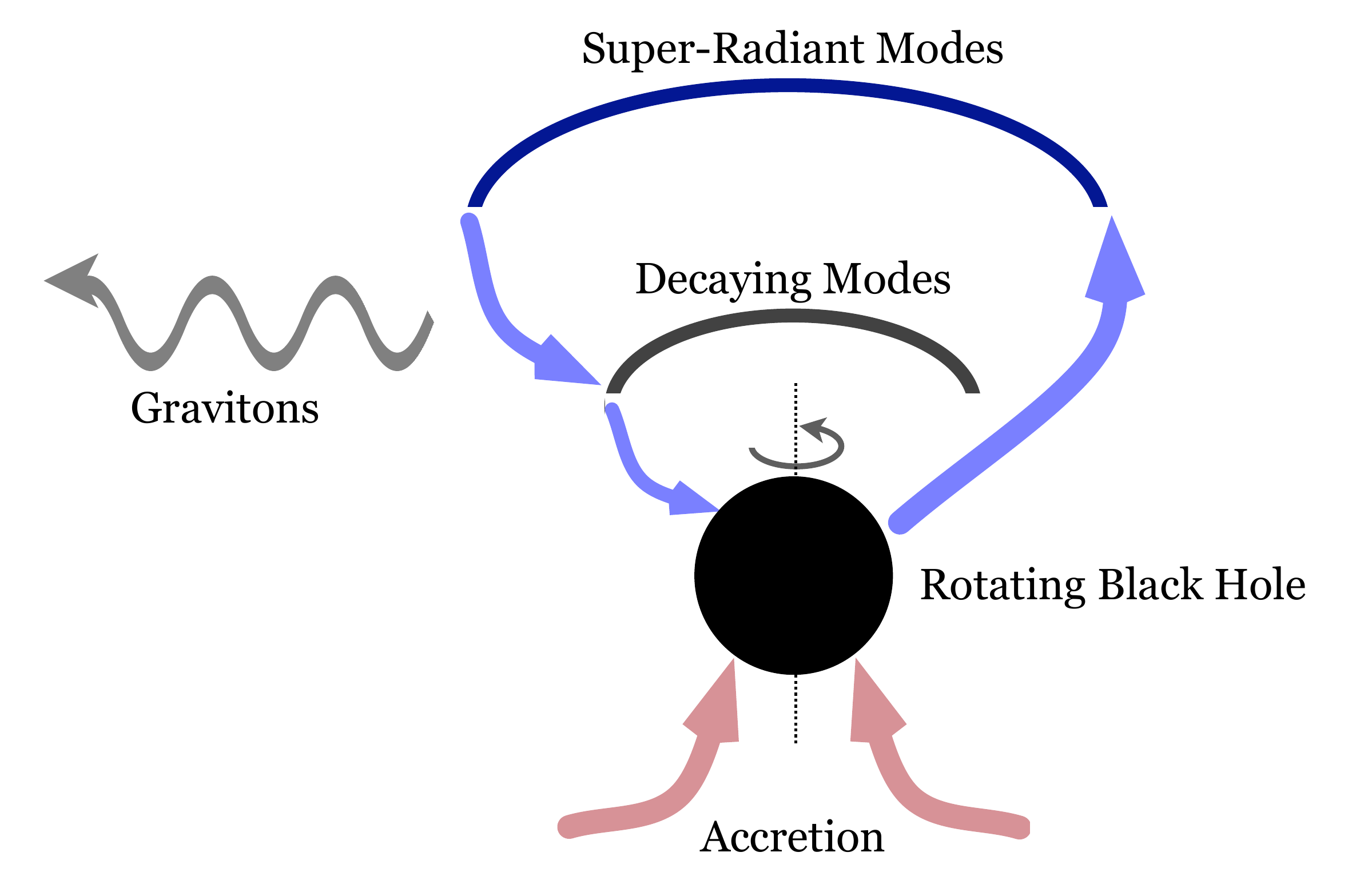}
 \caption{{\bf The Carnot Cycle of the Axionic Instability:} The black hole ``feeds" the superradiant state forming an axion Bose-Einstein condensate. Axions from that state quantum-mechanically transition through graviton emission to a lower-energy non-superradiant state. The non-superradiant state decays into the black hole. Accreting matter around the black hole replenishes the rotational energy lost to gravitons and sustains this cycle.}
 \label{carnotcycle}
 \end{center}
\end{figure}

A rotating black hole 
surrounded by an axionic cloud is essentially a huge quantum mechanical system similar to an atom, with a crucial difference  that particles populating its levels
are bosons, rather than fermions. As a result, some of the levels become highly populated. 
Moreover, a nucleus---the central black hole---continuously creates particles on some of the levels and destroys them on the others with the rate proportional to the occupation number of the level. A model independent source of back-reaction is related to the possibility for an axion in the growing superradiant cloud to emit a graviton and to jump onto a non-superradiant level, from where it eventually gets sucked inside the black hole horizon. 
The whole system is fueled by the inflow of  accreting matter, and, if the accretion is efficient enough, it acts as a giant gravitational
wave siren, see Figure~\ref{carnotcycle}. 

\begin{figure}[t!] 
  \begin{center}
  \includegraphics[width=5in]{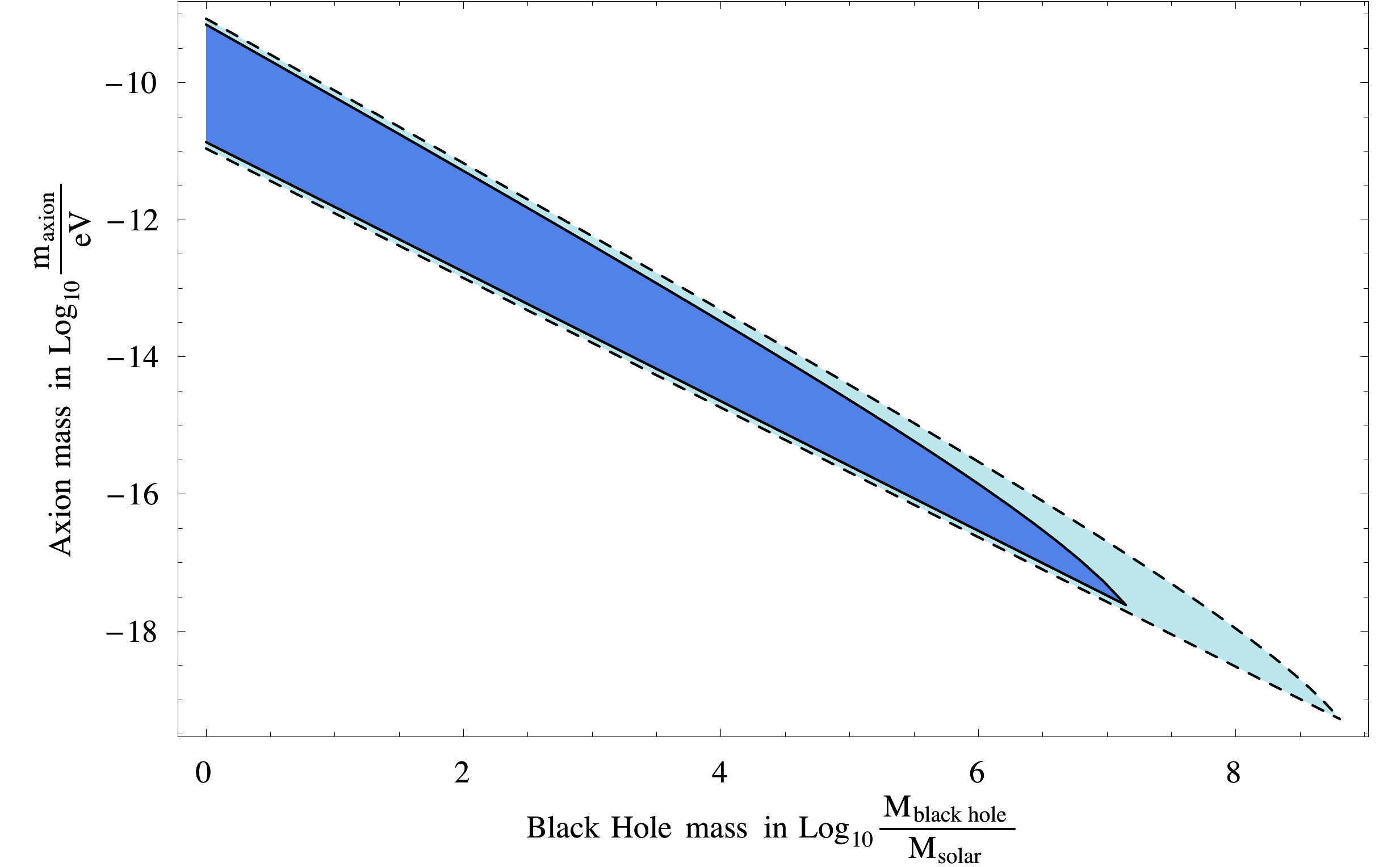}
  \caption{The parameter space in the plane of axion and black hole masses where the superradiance leads to the 
  black hole spindown assuming Eddington accretion  (dark shaded region), 
  and no accretion, {\it i.e.}  the superradiance time scale is required to be faster than the age of the Universe (light shaded area). In both cases the upper bound has been calculated from (\ref{heavy}) and the lower bound from (\ref{instimes}).}
  \label{bhbounds}
  \end{center}
\end{figure}

\begin{figure}[t] 
  \begin{center}
  \includegraphics[width=5in]{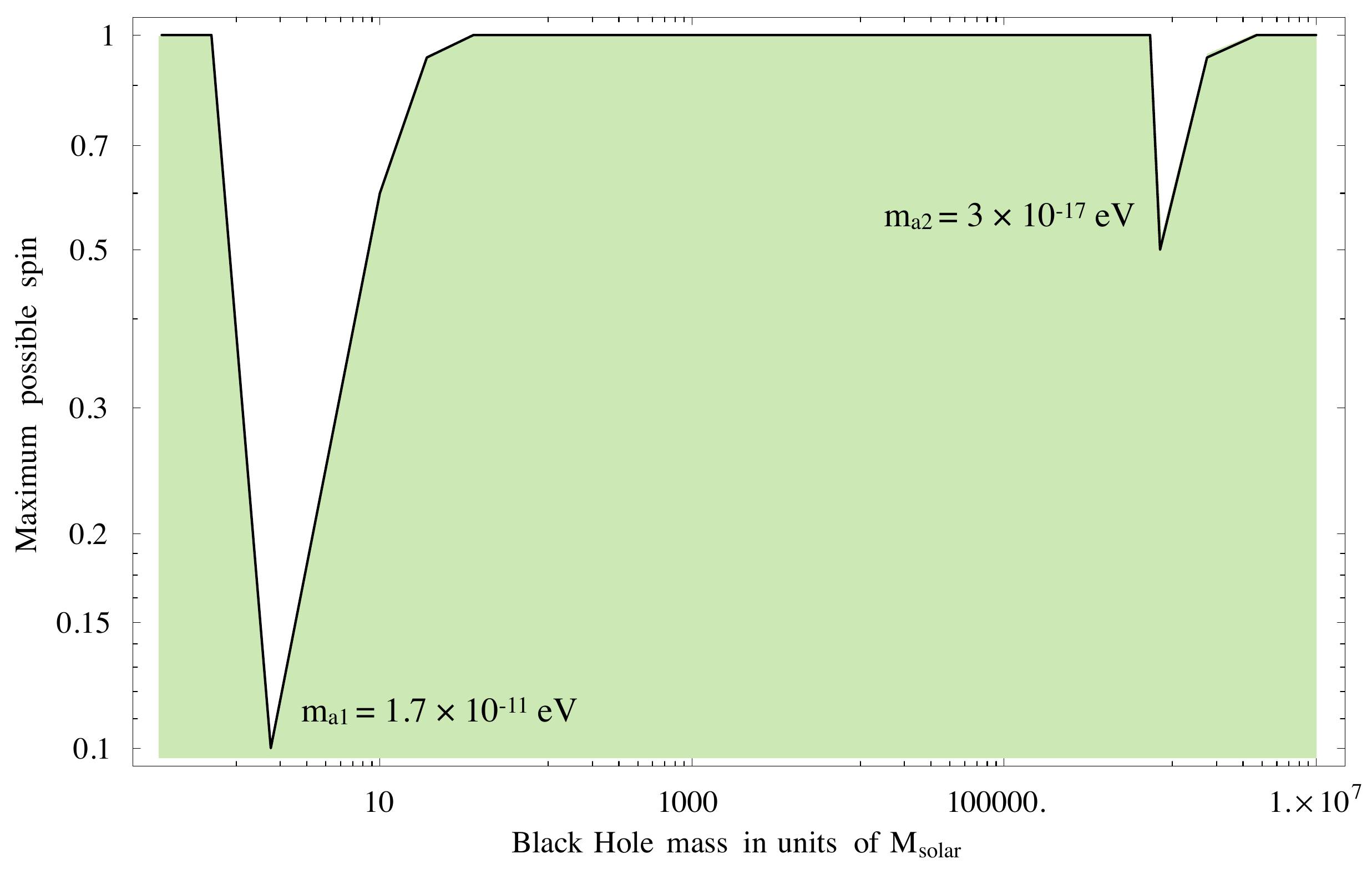}
  \caption{The maximum allowed spin for a black hole as a function of its mass assuming there are two axions with mass $m_{a1}$ and $m_{a2}$ corresponding roughly to black hole masses of $2M_\odot$ and $10^6M_\odot$. This plot has been created using (\ref{heavy}) (and the dependence of the superradiance rate on the black hole spin for heavy masses from \cite{Zouros:1979iw}) and (\ref{instimes}) and is indicative.}
  \label{maxspin}
  \end{center}
\end{figure}

 To gain some intuition about different regimes of the axionic siren it is instructive to write a simplified set of kinetic equations,
  describing the
evolution of the black hole, accretion disc and axionic cloud. 
Namely, let us characterize the siren by the total number of axions on the superradiant levels $N_+$, non-superradiant levels $N_-$, and the black hole
spin $J={a \over R_g}$. The evolution of the occupation numbers $N_+$, $N_-$ can be approximated by the following couple of kinetic equations,
\begin{gather}
\label{kin+}
{dN_+\over dt}=\tau_{+}^{-1}N_+-\tau_{GW}^{-1}N_+(N_-+1)\\
\label{kin-}
{dN_-\over dt}=-\tau_-^{-1}N_-+\tau_{GW}^{-1} N_+(N_-+1)
\end{gather}
Here, 
\be
\tau_\pm=\epsilon_\pm^{-1} R_g
\ee
are the superradiance and dumping time, and the graviton emission time $\tau_{GW}$ is
\be\label{gwemission}
\tau_{GW}\simeq M_{Pl}^2R_g^3\;,
\ee
assuming the axion mass is close to the optimal value, $\mu\sim R_g^{-1}$. This two-level approximation becomes literally true in the low mass limit
$\mu\lesssim R_g^{-1}$, when the axion cloud is dominated by $2p$ states. The pair of kinetic equations (\ref{kin+}), (\ref{kin-}) has a stationary point, at which
\be
\label{stationary}
N_\pm={\tau_{GW}\over \tau_\mp}\simeq \epsilon_\mp M^2_{Pl} R_g^2= \epsilon_\mp {M R_g}\;.
\ee
As we expected, in this stationary regime the total mass of the axion cloud $M_a\simeq\mu (N_++N_-)$ is small compared to the black hole mass.
In order to establish this regime, the accretion of matter into the black hole should be efficient enough, otherwise the black hole would simply loose its
spin and the siren would not start. To see when the stationary regime (\ref{stationary}) can be established for a rapidly rotating black hole, let's 
consider the following  kinetic equation describing the evolution of the black hole spin,
\be
\label{spineq}
{dJ\over dt}=\tau_{accr}^{-1}J-{\dot J}_{GW}\;,
\ee
where we assumed the Eddington regime for the spin accretion, so that the accretion time is given by (\ref{eddtime}).
The spindown rate due to gravitational wave emission ${\dot J}_{GW}$ at the stationary point (\ref{stationary}) is determined by the number of 
gravitons emitted per unit time (each graviton carries away two units of the angular momentum)  
\be
\label{spindown}
{\dot J}_{GW}\approx2\tau_{GW}^{-1}N_+N_-\simeq MR_g{\epsilon_+\epsilon_-\over R_g}=\epsilon_-\tau_{sr}^{-1}\;.
\ee 
Consequently, the axionic siren can operate in a stationary regime provided 
\be
\label{sirenregime}
\tau_{sr}>\epsilon_-\tau_{accr}\;.
\ee
Note, that this condition is different from the naive one $\tau_{sr}>\tau_{accr}$, the reason being that the axion cloud is small, so that less accretion is needed to support it. In the opposite regime, $\tau_{sr}<\epsilon_-\tau_{accr}$ the accretion is not efficient enough to support
the siren, and the black hole spins down. Consequently, in the presence of light axions 
we expect to see gaps in the spectrum of rotating black holes at the masses close to the optimal. We illustrated this effect in Figures~\ref{bhbounds},
\ref{maxspin}. 
In Figure~\ref{bhbounds} we  show the regions in the black hole mass/axion mass parameter space, where the condition (\ref{sirenregime})
cannot be satisfied for a maximally spinning black hole and, consequently, black holes cannot sustain maximal rotation.  The dark shaded region arises for Eddington limited accretion, see (\ref{eddtime}), while the light shaded region for an accretion time of order the age of the Universe, i.e. currently non-accreting black holes.
In this plot we assumed that $\epsilon_-=\epsilon_+$ and that $\tau_{sr}$ given by (\ref{heavy}) and (\ref{instimes}) for the heavy and light axion mass regime, respectively. This plot should be considered only as an indicative one and the more
refined analysis will be presented in \cite{inprogress}. In particular, the estimate (\ref{gwemission}) for the graviton emission time should be considered as a lower bound, as it does not take into account non-relativstic suppression. Taking this suppression into account will result in the stronger bounds on the axion masses than the conservative estimates presented here.
We checked, however, that setting $\epsilon_-=1$ in (\ref{sirenregime}) affects the allowed axion masses in the heavy region only by a factor of order two
(while the lower boundary of the spindown region in  Figure~\ref{bhbounds} is more sensitive to such a change).

In Figure~\ref{maxspin} we illustrated how the black hole Regge plane (the parameter space of black hole masses and spins) may look like
after 
precision measurements of spins and masses for many black holes will become available. Namely, in the presence of two axions with masses $1.7\times10^{-11}$eV and $3\times10^{-17}$eV black holes will populate only the shaded region in this plot (there could be rare exceptions due to black holes experiencing a short period of superEddington accretion). This figure should be considered as indicative, and 
is meant to illustrate that the dip in the spectrum of rotating black holes becomes narrower and less pronounced with increasing black hole mass.

Heavy enough black holes may satisfy the condition (\ref{sirenregime}) and operate in the siren regime. The graviton flux (\ref{spindown})
corresponds to the gravitational wave signal at the Earth of the strength
\be
\omega h^2\sim {\dot{J}_{GW}\over M_{Pl}^2 R_g^2}\l{R_g\over L}\r^2={\epsilon_+\epsilon_-R_g\over L^2}\;,
\ee  
where $L$ is the distance to the siren.
For $\epsilon_\pm=10^{-7}$ this translates into
\be
h\sim3\times 10^{-22}\l 10^{-2}{\rm Hz}\over \nu\r^{1/2}\l{ M\over 10^7 M_\odot}\r^{1/2}\l100{\rm ~Mpc}\over L\r\;, 
\ee
which is above LISA sensitivity.

We should stress, that the stationary point (\ref{stationary}) is the simplest possible regime for the axionic siren. However,
the siren is a very rich and complicated dynamical system that may exhibit other even more colorful periods during its life-time.
For instance, if we deviate from the stationary point (\ref{stationary}) of equations (\ref{kin+}) and (\ref{kin-}), we find cyclic solutions with the axion population oscillating between the superradiant and the non-superradiant level.
 These solutions would give rise to a gravitational pulsar with a period of order the superradiance time (10 years for a $10^7M_\odot$ black hole), 
 with a maximum amount of radiation during the periods when the population of both levels are comparable.
 A cyclicity of such a pulsar is a direct manifestation of the quantum origin of the
 axionic siren.

We should mention that the above discussion ignores a number of important physical effects, such as the back reaction from self-interactions in the axion cloud, and one graviton annihilation of two axions in the same level, but they do not change the qualitative picture presented above. These effects will be discussed in an upcoming paper \cite{inprogress}.
\subsubsection{Superradiance and the QCD Axion}
\label{axion_bhbounds}

For $f_a\sim M_{pl}$ the QCD axion mass is $5 \times 10^{-13}$ eV; given that the smallest stellar size black holes may have masses down to $\sim 2M_\odot$, their
 spin measurements may probe axion masses up to $3 \times 10^{-10}$ eV, exploring the 
 parameter space of the QCD axion deep in the anthropic regime. In terms of the QCD axion decay constant
 this translates into
\be
f_a > 2 \times 10^{16}~\mbox{GeV},
\ee
{\it i.e.} down to the GUT scale---the natural scale for the PQ symmetry breaking. Note that this reach is in a region where all electromagnetic couplings are suppressed and is also independent of the axion's cosmological abundance. From the already existing spinning black hole candidates, 
the best bound is provided by
LMC X-1---a 10$M_\odot$ black hole with spin ${a / R_g}\approx0.91$ \cite{Gou:2009ks}.
These values suggest a bound for  $f_a$ at the level 
\[
f_a\lesssim2 \times 10^{17}\mbox{ GeV}\;,
\]
which is significantly below the Planck scale.  

Again, both these numbers should be considered as indicative (and conservative) estimates. Our preliminary results including the effects mentioned in the previous section and not accounted
here, indicate that the actual bounds are likely to be stronger by a factor $\sim 2$.

To conclude, it is worth noting that one could use the superradiance effect  to place bounds on the photon and graviton masses.
 In the photon case, though, an electromagnetic field  resulting from the instability  will interact strongly with the matter in the black hole accretion disk and is most likely  to dissipate. For the graviton, only the largest galactic black holes have a chance to compete with a limit from the binary pulsar timing.
 It was already suggested that these can be used to probe  massive gravity models by gravitational wave measurements sensitive to the presence of black hole hair \cite{Dubovsky:2007zi};
 in order to exploit  the superradiance effect it is necessary to observationally confirm the presence of an ergosphere region, which is not automatic in massive gravity, where the Kerr
 metric gets modified.  An interesting consequence of a non-vanishing graviton mass would be that in the heavy mass regime graviton emission becomes
 impossible, and does not prevent the graviton cloud from becoming an order one perturbation.
 
 \section{Discussion and Future Directions}
\label{discussion}
We see that taking seriously the QCD axion as a solution to the strong CP problem together with generic properties
of axions in string theory leads to a rather unconventional set of predictions for forthcoming cosmological and astronomical
observations.
 Of course, it would be exciting to observe any of the signatures discussed in the current paper---rotation of the CMB polarization, a step in the CDM power spectrum or a gap in the
mass spectrum of rotating black holes. However, a really distinctive property of the presented scenario  is the expectation of {\it many} light axion fields.
This multiplicity can reveal itself by giving rise to all of the above signals simultaneously. Even more interestingly, precision measurements of the CDM power spectrum and   spectroscopy/gravity wave signals from  rotating black holes are capable not only of providing the evidence for the presence of many light axions but also to measure values
for many parameters of the same physical origin, such as axion masses/PQ breaking scales, and initial misalignments angles.
With large enough number of axions this may allow to make quantitative tests of different scenarios 
 for the statistical distribution of these parameters in the landscape and/or of inflationary measures.

As discussed in the {\it Introduction}, if many ultra-light scalar fields were to be observed, it is tempting to suggest that they may be responsible for the scanning of the vacuum energy---at least at low scales. Indeed, in the current framework there are at least three natural sources for the scanning. First, the vacuum energy can be scanned at the high scale by fluxes, as in the 
Bousso--Polchinski scenario. Second, we expect a large number of real scalar moduli $\phi_i$; their potential is generated as a result of SUSY breaking and depends on the SUSY breaking vev
$F$ and the string scale $M_s$ in the following way,
\[
V(\phi_i)=F_{susy}^2f(\phi_i/M_s)\;.
\]
 This potential may provide a possibility to scan the vacuum energy below the SUSY breaking scale. Note, that unlike in the toy field theory landscape of \cite{ArkaniHamed:2005yv}, the 
 maximum scale of scanning in this case is parametrically above the mass scale $F/M_s$ for these fields. 
 Finally, the scanning can be done by axions. In this case the masses of the fields are also parametrically below the overall scale of the potential. The difference with the scanning by real moduli is that the axion masses are exponentially sensitive to the parameters of compactification, and likely to be distributed over many orders of magnitude rather than being concentrated around one particular scale, giving rise to the situation close to the Wilsonian scanning.
 
 The latter possibility raises a number of theoretical and phenomenological issues. First, as we already discussed, the scanning is only possible, provided individual axion potentials have non-degenerate minima. This implies that axions responsible for the scanning are associated with a gauge sector strongly coupled at low energies. Note, that by itself this does not  guarantee
 the presence of multiple axion vacua. For instance, the axion potential in pure gluodynamics is quadratic in the large $N$ limit \cite{Witten:1998uka}, so that there is a single axion vacuum. 
 
 Note that in this example
 the strongly coupled sector itself, even in the absence of the axion, possess $N$ non-degenerate metastable vacua. These vacua are useless for the scanning, however.
  First, not all of them are extrema of the axion potential. Moreover, even if in a more complicated setup, this kind of vacua become local minima of the axion potential, they still do
  not allow to scan the vacuum energy if the strong coupling scale $\Lambda$ of the gauge sector is lower then the expansion rate of the Universe during inflation, $H_{inf}$. Indeed, at the moment when at the FRW stage the expansion rate drops below $\Lambda$, the gauge sector experiences the (de)confinement phase transition (assuming it was not reheated, otherwise it happens later) and a network
  of domain walls separating different vacua is formed. Afterwards,  bubbles of the lowest energy vacuum  expand at the expense of the other vacua and the Universe always
  ends up being in the lowest energy state. This has to be contrasted with the axion vacua that exist only in the presence of a dynamical axion. Those are separated by a large distance of
  order $f_a$ in the field space. So if $H_{inf}\ll f_a$  (but, possibly,   $H_{inf}\gg\Lambda$) the axion fluctuations are negligible compared to the distance between vacua and the
  network of domain walls does not  form.
 We see that an important theoretical issue that has to be understood is what are the conditions for developing
 non-degenerate vacua in the axion potential and how generic such a situation is. 

If axions scan a significant fraction of the vacuum energy in the Wilsonian way, there should be several hidden gauge sectors
per decade of energy at low energy scales. How is it possible that they avoided being detected so far? The natural answer to this question would be that these sectors
are well separated from us along extra dimensions, so that locality sequesters them away. For instance, if inflation were due to some well-localized 
process in extra-dimensions (such as the brane inflation) 
this would explain why only the visible sector was reheated.

On the other hand, for some purposes a large separation along extra dimensions may not be enough to realize sequestering at the desired level. For instance, it is conceivable that some
of the hidden sectors posses massless $U(1)$ gauge factors and have light  (with masses smaller than $\sim 10$~keV) fermions charged under these gauge bosons. In this case, there are
extremely strong 
 bounds from star cooling on the kinetic mixing coefficient $\epsilon$ between the photon and an extra massless gauge boson $\epsilon\lesssim 10^{-14}$
 \cite{Davidson:2000hf}. On the other hand, it was argued \cite{Abel:2008ai} that even if the extra $U(1)$ is separated from our sector by a large distance in string units, this mixing can be mediated by light (compared to the string scale) closed string modes at the unacceptable level.
 Consequently, the smallness of such kinetic mixings requires a separate explanation.  For instance, it might be a consequence of a large mass  for the closed string modes capable
 of mediating such a mixing. It would be enough if such a mass were generated just in the vicinity of where the photon is localized. 
 
 Note, that similar to the logic in the {\it Introduction} that led us to the expectation that {\it many} light axions may be present, one can also make the case for  extra hidden gauge sectors\footnote{
 One difference is that unlike for the QCD axion it is possible to argue that the existence of at least one gauge sector is required anthropically.}.
 Consequently, these phenomenological issues remain even if axions are not responsible for the scanning.
 
 Another feature that can be generic in the compactification manifold is the existence of warped throats \cite{Verlinde:1999fy,Dimopoulos:2001ui,Hebecker:2006bn} which generate hierarchically small mass scales by the Randall-Sundrum mechanism \cite{Randall:1999ee}. Warping can naturally lower the value of $f_a$ below $M_{GUT}$ -- $f_a$ can now vary over a large range of mass scales, affecting the phenomenology of these axions. For example, the predicted suppression in matter perturbations due to light axions changes, or, as explained in Section \ref{axion_bhbounds}, the effect of superradiance can be inhibited due to the larger possible coupling to $\vec E \cdot \vec B$. In addition, the scale of gauge sectors in warped throats can naturally be well below the string scale and may be giving rise to new ultra-light fields. These ultra-light fields can also include higher spin excitations gravitationally coupled to us. 
  The phenomenology of axions with varying $f_a$ below the GUT scale, as well as possible implications from the existence of ultra light hidden sectors will be the subject of a companion paper.

\section{Acknowledgements}
We thank T.~Abel, M.~Aganagic, R.~Blandford, D.~Green, K.~Jedamzik, S.~Kachru,  R.~Romani, E.~Silverstein, N.~Toro, H.~Verlinde, B.~Wagoner, A.~Westphal and M.~Zaldarriaga for extremely helpful and enjoyable discussions. The work of NK is supported in part by the DOE Grant DE-FG03-91ER40674. 
JMR gratefully thanks the UC Berkeley Center for Theoretical Physics for their hospitality during the course of
this work.  JMR is partially supported by the EC network 6th Framework Programme Research and Training Network Quest for Unification (MRTN-CT-2004-503369), by the EU FP6 Marie Curie Research and Training Network UniverseNet (MPRN-CT-2006-035863), and by the STFC (UK). 
  
\section{Appendix: Superradiance as Anti-tunneling}
\label{tunneling}

\begin{figure}[t] 
 \begin{center}
 \includegraphics[width=5in]{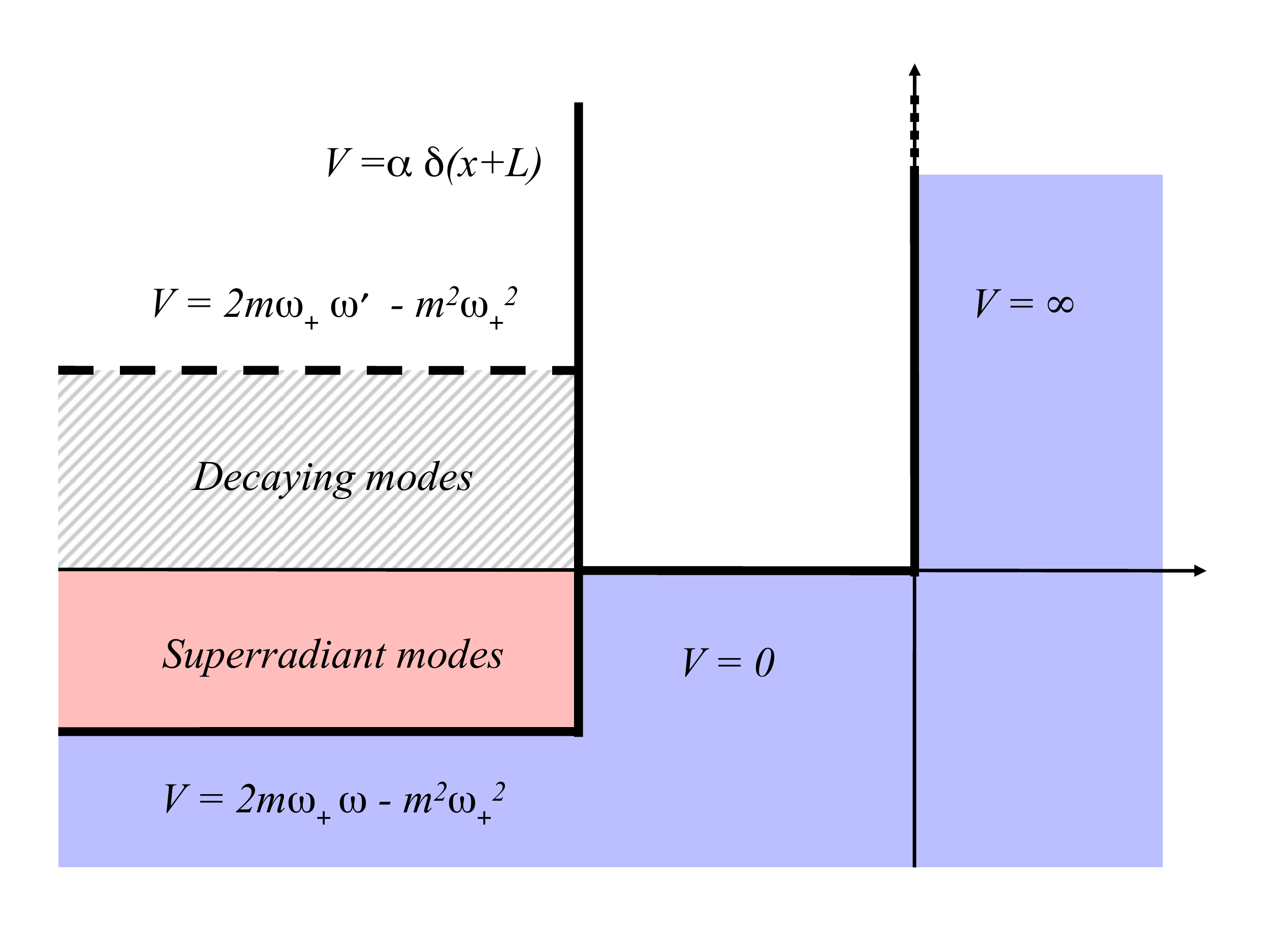}
\caption{A toy model potential encapsulating all the salient features of the superradiant instability: 
infinite potential barrier, representing the mass mirror; a potential well with $V=0$, a reasonable approximation for the modes near its top; a strongly repulsive Dirac $\delta$-function potential simulating the centrifugal barrier; and a potential $V = 2 m \omega_+ \omega - m^2 \omega_+^2$,  modeling the dispersive properties of the waves far in the ergoregion as they approach the horizon, where they satisfy $k^2 = (\omega - m \omega_+)^2$, such that the net potential there is large and positive for large eigenvalues, ($\omega' $ on the figure), but is negative for superradiant modes (denoted by 
$\omega$ on the figure). The key is that the superradiant modes have larger momentum outside the well than inside it.}
 \label{potentialtunnel}

\end{center}

\end{figure}

In Section \ref{BHbomb} we have introduced the description of the exponential superradiant instability
as a tunneling phenomenon, based on the Schr\"odinger problem (\ref{schrod}), (\ref{schrodpot}). Here we will discuss a simplified version of the problem which clearly demonstrates how the superradiance sets in. Let us for this purpose consider the $1D$ Schr\"odinger problem
\be
\Psi'' + \Bigl(\omega^2 - V(x,\omega)\Bigr) \Psi = 0 \, ,
\label{schreq}
\ee
where the potential is given by (see Figure \ref{potentialtunnel})
\be
V(x,\omega) = \begin{cases}  \infty \, , &~~~ x > 0 \, ; \cr
 \alpha \delta(x+L) - \Bigl(2m \omega_+ \omega - m^2 \omega_+^2 \Bigr) \Bigl(1- \Theta(x+L)\Bigr) \, & ~~~ x<0\, . 
 \end{cases}
\label{toypotential} 
\ee

\noindent The potential (\ref{toypotential}) faithfully represents (\ref{schrodpot}); it has: a mirror at large distances from the black holes, here modeled by the infinite potential barrier at the origin; the centrifugal potential at distances of the order of black hole's gravitational radius, represented by the strongly repulsive $\delta$-function at $x = -L$, with $\alpha/\omega \gg 1$; the potential well in between, with a negligible influence on the modes near the top of the well, reflected in our choice of $V=0$ in the well of (\ref{toypotential}); and the potential $V = 2 m \omega_+ \omega - m^2 \omega_+^2$ on the horizon's side of the $\delta$-barrier, which is chosen to encode the dispersion relation of a wave with angular momentum $m$ in the near-horizon limit, where the wavevectors obey $k^2 = (\omega - m \omega_+)^2$. Note, that albeit this problem is straightforward to solve, it is {\it not} of the usual self-adjoint form as commonly encountered in quantum mechanics on Hilbert space. Instead, the potential itself depends on the eigenvalue, which as we will now see has dramatic consequences for the existence of the instability. 

Let us now solve this equation. Away from the barriers, the solutions are given by linear combinations of
free waves $e^{\pm i q x}$, with $q = \sqrt{\omega^2 - V}$ in the corresponding region. At $x=0$, where the wavefunction enters the barrier, we set $\Psi = 0$ since the barrier is infinite both in height and in width, and penetration depth therefore vanishes. Far to the left of the $\delta$-barrier, as we noted in Section \ref{BHbomb}, the regularity at the horizon picks the wave with the momentum 
dependence $\exp( -i k_+ x)$ as the correct asymptotic form of  the eigenmodes of (\ref{schrod}), 
(\ref{schrodpot}) \cite{Zouros:1979iw}, where $k_+$ is formally the positive root of 
$q^2 = \omega^2 - V = (\omega-m \omega_+)^2$; thus $k_+ = \omega - m \omega_+$. Note, that when 
$\omega < m \omega_+$, this root is {\it negative}:
it describes a wave which is incident on the $\delta$-barrier from the {\it left}. Nevertheless, the group velocity of this wave is still positive, $v_g = \frac{d\omega}{dk} = 1$, implying that any wavepacket composed of such modes moves away from the $\delta$-barrier, even if the phase velocity of the wave changed sign. This is of course the first hint of the superradiant behavior in our simple toy model problem: the waves carrying away {\it negative} energy to the left leak into the potential well on the right, and amplify the state bound up in there. In fact, for the modes $0 < \omega < m\omega_+$ we see that this boundary condition really represents the exact opposite of the usual textbook examples of tunneling, since the `free' wave impinges into the barrier, and so we can dub it `anti-tunneling'. 

What remains is to complete the determination of the spectrum of (\ref{schreq}), (\ref{toypotential}) by matching the waves in the well and out in the ergoregion across the $\delta$-barrier. There, the solutions obey
\be
\Psi_L|_{x=-L} = \Psi_R|_{x=-L} \, , ~~~~~~~~ \Psi_R'|_{x=-L} - \Psi'_L|_{x=-L} = \alpha \Psi|_{x=-L} \, .
\label{deltabcs}
\ee
Substituting
\be
\Psi_L = B e^{-i(\omega - m \omega_+) x} \, , ~~~~~~~~~~~~~~~ \Psi_R = A \sin(\omega x) \, ,
\label{waves}
\ee
where these functions are picked to satisfy the boundary conditions far to the left and at the origin, respectively, we get a secular equation for the eigenvalues $\omega$:
\be
\omega \cot(\omega L) + \alpha = i \Bigl(\omega - m \omega_+ \Bigr) \, .
\label{eigeneqs}
\ee
Two comments are in order here. First: the secular equation (\ref{eigeneqs}) should come as no surprise, as we are dealing with a problem that involves three modes and four boundary conditions, thanks to the horizon ingoing condition and infinite barrier cutoff enforcing (\ref{waves}), and (\ref{deltabcs}) relating their parameters at the $\delta$-barrier. Physically, this merely means that the potential well of Figure \ref{potentialtunnel} can only accommodate states which have right phases to fit inside it, as it enforces two-sided boundary constraints. This is expected, as any bound state spectrum is indeed discrete. Second: the eigenfrequencies all have imaginary parts. This is because the $\delta$-barrier is not impenetrable, unlike the infinite barrier at the origin, but allows leakage which links the bound states with the continuum of outgoing waves to the left. This is indeed familiar from the usual tunneling problems, and reflects the fact that under time evolution the bound states can evolve into free states that escape to infinity. In
the standard tunneling problems, however, the imaginary parts of the eigenfrequencies are all negative, which means that the bound states decay as time goes on. That is again a natural consequence of a setup which one adopts, which is that a quantum state is prepared in a surrounding vacuum and allowed to fall apart by the emission of a wave through a partially transparent barrier. 

The unusual feature of (\ref{eigeneqs}) is that many of the eigenmodes have {\it positive} imaginary parts. This can be readily seen by setting 
\be
\omega = \omega_r - i \Gamma \, , 
\label{defeq}
\ee
and solving (\ref{eigeneqs}) in the limit $\omega L \gtrsim 1$ (which means that the well has many closely populated bound states, but that we look for ones near the top) and $\alpha/\omega \gg 1$ (so  that the $\delta$-barrier is strongly repulsive, and the bound states are long-lived, allowing for an adiabatic approach to the problem). These approximations directly follow from the black hole instability considerations, which we are principally interested in. Then, the secular equation (\ref{eigeneqs}) splits into two,
\bea
&&\omega_r \cot(\omega_r L) + \alpha + {\cal O}(\Gamma) = 0 \, , 
\label{freqs} \\
&& \Gamma \cot(\omega_r L) = - (\omega_r - m \omega_+) + {\cal O}(\Gamma \omega_r L) \, .
\label{eigens}
\eea
When $\alpha/\omega \gg 1$ the standard magic of cotangents in the tunneling 
eigenequations comes to the rescue: the eigenmode real parts are determined by the limiting 
form of the equation (\ref{freqs}),
$\cot(\omega_r L) \simeq - \alpha/\omega_r$, and are approximately 
\be
\omega_r \sim \frac{n \pi}{2L} + \ldots \, .
\label{energies}
\ee
This implies that the solutions of the other secular equation (\ref{eigens}) are
\be
\Gamma \simeq \frac{\omega_r (\omega_r - m \omega_+)}{\alpha} + \ldots \, .
\label{widths}
\ee

This is the key equation for understanding the superradiant instability. We see immediately that the modes in the regime $0 < \omega_r < m\omega_+$ have $\Gamma <0$, which from Eq. (\ref{defeq}) implies that these modes have {\it positive} imaginary contributions to the eigenfrequency. This means that, as time goes on, these modes {\it grow} exponentially. They are unstable, absorbing energy and spin from the black hole, rather than the other way around. For them, indeed, the `transmitted' wave is really incident as $k_+ = \omega - m \omega_+ <0$, and so they anti-tunnel, as we anticipated above. Finally, we also see that the instability scale $\Gamma$ peaks at $\omega_r \simeq m \omega_+/2$, as that is the extremal value of $\Gamma$ in Eq. (\ref{widths}). This of course presumes that the $\delta$-barrier is independent of the eigenfrequency $\omega$, as well as that the mirror barrier is fully impenetrable, as modeled by our infinite barrier at the origin. In the real black hole problem, neither is true. The real centrifugal barrier depends on the eigenmode, through the respective value of $l$. And the mirror is not completely impenetrable, being eventually completely transparent to the modes with $\omega > \mu$. All of this will modify the precise formulas for the superradiant eigenfrequencies and the instability time scales. 
Nonetheless,  we believe the above toy model is instructive and  accurately reproduces the essential features of the instability.

\end{document}